\documentclass[runningheads]{llncs}
\usepackage[T1]{fontenc}
\usepackage{graphicx}
\usepackage{times}
\usepackage{latexsym}

\usepackage[T1]{fontenc}
\usepackage[utf8]{inputenc}

\usepackage{microtype}

\usepackage{inconsolata}

\usepackage{amsmath,amssymb,amsfonts}
\usepackage{algorithmic}
\usepackage{textcomp}
\usepackage{xcolor}
\usepackage{tabularx}
\usepackage{multirow}
\usepackage{enumitem}
\usepackage{graphicx} % Required for including images
\usepackage{subcaption} % Required for subfigures
\usepackage{booktabs}
\usepackage{xcolor}
\usepackage{mdframed}
\usepackage{hyperref} % Required for \url and \urldef
\usepackage{caption} % Required for captions
\usepackage{array}
\usepackage{makecell}

\AtBeginDocument{%
	}
\usepackage{rotating}
\usepackage{tikz}
\def\checkmark{\tikz\fill[scale=0.4](0,.35) -- (.25,0) -- (1,.7) -- (.25,.15) -- cycle;} 

\begin{document}
\title{An Investigation of Prompt Variations\\for Zero-shot LLM-based Rankers}

\author{Shuoqi Sun\inst{1}\thanks{This work was conducted while Shuoqi Sun was a student at The University of Queensland.}\orcidID{0009-0000-9329-9731} , Shengyao Zhuang\inst{2}\orcidID{0000-0002-6711-0955}, Shuai Wang\inst{3}\orcidID{0000-0002-0726-5250}, Guido Zuccon\inst{3}\orcidID{0000-0003-0271-5563} 
}
%\authorrunning{S. Sun et al.}

\institute{
RMIT University, Melbourne, Australia \\
\email{shuoqi.sun@student.rmit.edu.au} \and CSIRO, Herston, Australia \\ \email{shengyao.zhuang@csiro.au} \and The University of Queensland, St. Lucia, Australia \\ \email{\{shuai.wang2, g.zuccon\}@uq.edu.au}}
\maketitle    

% --- PUBLICATION NOTICE BOX ---
\thispagestyle{empty} % Optional: removes page number from the first page
\begin{center}
\fcolorbox{black}{gray!15}{%
  \begin{minipage}{0.95\textwidth}
    \textbf{Publication Notice:} This paper has been accepted for publication at the 47th European Conference on Information Retrieval (ECIR 2025). The \href{https://link.springer.com/chapter/10.1007/978-3-031-88711-6_12}{published version} is available in the conference proceedings.
  \end{minipage}%
}
\end{center}

% typeset the header of the contribution
%
\begin{abstract}
	We provide a systematic understanding of the impact of specific components and wordings used in prompts on the effectiveness of rankers based on zero-shot Large Language Models (LLMs).
	Several zero-shot ranking methods based on LLMs have recently been proposed. Among many aspects, methods differ across (1) the ranking algorithm they implement, e.g., pointwise vs. listwise, (2) the backbone LLMs used, e.g., GPT3.5 vs. FLAN-T5, (3) the components and wording used in prompts, e.g., the use or not of role-definition (role-playing) and the actual words used to express this. 
	It is currently unclear whether performance differences are due to the underlying ranking algorithm, or because of spurious factors such as better choice of words used in prompts. This confusion risks to undermine future research.
	Through our large-scale experimentation and analysis, we find that ranking algorithms do contribute to differences between methods for zero-shot LLM ranking. However, so do the LLM backbones -- but even more importantly, 
	the choice of prompt components and wordings affect the ranking. In fact, in our experiments, we find that, at times, these latter elements have more impact on the ranker's effectiveness than the actual ranking algorithms, and that differences among ranking methods become more blurred when prompt variations are considered.  
	
	%These are important findings because they (1) clarify whether improvements offered by specific methods should be attributed to the underlying ranking algorithm or less understood components related to the prompts, (2) inform search engine practitioners on the aspects that should be compared and evaluated with attention in the context of zero-shot LLM-based rankers.
	
	\vspace{-6pt}
\end{abstract}

\sloppy
\section{Introduction}
\vspace{-6pt}

Large Language Models (LLMs) are massively parametrised transformer models that have undergone training on a large extent of text~\cite{zhao2023survey}. Generative LLMs are capable of generating text in response to textual input (the prompt or context)~\cite{brown2020language}. This prompting facility has been used to instruct LLMs to perform specific tasks~\cite{brown2020language,wang2023can,white2023prompt,zhuang2024setwise,fan2023recommender,zhuang2024promptreps}. 
%,fan2023recommender

In this paper, we investigate the use of LLMs to create zero-shot  rankers %\footnote{We specifically focus on re-rankers, where an initial set of documents is retrieved from an index using a first-stage retriever, and a subset is provided to the re-ranker to re-order. For ease of reading, we use rankers, in place of re-rankers throughout the paper.}
\cite{ma2023zero,sun2023chatgpt,tang2023found,zhuang2024beyond,qin2023large,zhuang2024setwise}. With zero-shot, we mean ranking methods that are not specifically trained for ranking tasks (i.e. not contrastive fine-tuning). These rankers operate by following the instructions provided in the prompt, which include the query  and the $k$ documents that should be considered for ranking. Using the LLM, rankers then generate an answer to comply with the ranking instruction. Finally, either the generated answer contains the ranking provided by the method, or the logits of the answer are examined to infer a ranking. % (or part of the ranking, and this operation is repeated to form a full ranking). 

Four families of zero-shot LLM rankers have been proposed: pointwise~\cite{zhuang2024beyond}, pairwise~\cite{qin2023large}, listwise~\cite{ma2023zero,sun2023chatgpt,tang2023found}, and setwise~\cite{zhuang2024setwise}. They differ because of the ranking algorithm (or mechanism) implemented in the instructions described in the prompt. For example, in pointwise the LLM is instructed to determine the relevance of a document, while in pairwise the LLM is instructed to determine which of two documents is more relevant. 
Within each family, one or more methods have been proposed. They typically differ in the backbone LLM, the wording of the prompts, and in how the ranking is formed once the answer is generated by the LLM. %, e.g., using the model logits or the actual content of the generated text. 

Recent work has shown that, once these zero-shot LLM rankers are compared using the same backbone LLM and fixing how the ranking is derived (i.e. generation vs. logits), setwise methods are the most effective~\cite{zhuang2024setwise}. Depending on the dataset, pairwise and listwise methods are similarly effective, with pointwise methods providing lower effectiveness overall. 
While they did recognise that the use of different backbone LLMs in previous work biased the comparison between methods, they did not identify that the actual prompts used by the different rankers differed not just in terms of the words used to describe the ranking algorithm, i.e. the instruction associated with how scoring should be performed, but also in terms of ``\textit{accessory wording}'' used. For example, compare the prompt of the PRP pointwise approach~\cite{qin2023large} and the one of the RankGPT listwise approach~\cite{sun2023chatgpt} -- see Table~\ref{prompt-wording-example}: RankGPT's prompt includes also a ``role-playing'' component (in the system part of the prompt: ``You are RankGPT [...]''), absent in PRP.

With this respect, then, we ask: \textit{What is the effect of such differences in wording used in the prompts?} And, more broadly: Are differences in effectiveness due to the actual ranking algorithm, or they are due to the choice of words used in the prompts? Are differences due to LLM characteristics such as backbone and size?  These are important questions because answering them will give us an understanding of what impacts the effectiveness of LLM rankers%(i.e., ranking methodology, backbone LLM, use of generated text vs. logits, accessory wordings used in the prompt)
, and will influence how methods should be compared in future.
We explore these directions of enquiry through a wide array of experiments, for a total of 1,248 prompts, in which we fix the backbone LLMs of the rankers, and we vary prompts in a controlled manner, categorising different prompt components and investigating their effects.

%\todo{give an highlight of interesting results}
%
%
%\todo{give a bullet point summary of contributions}

\vspace{-12pt}
\section{Related Work}
\vspace{-8pt}

\subsubsection{Sensibility of LLMs to Prompt.}
Previous work has shown that LLMs are sensitive to the prompt formulation~\cite{kim2023better,thomas2024large,koopman2023dr,kamruzzaman2024prompting,anagnostidis2024how-susceptible,mizrahi2024state}.
For example, Kim et al.~\cite{kim2023better} explored the effect of various prompts and prompting technology on  deploying LLMs across language generation tasks, showing dependency between prompt and effectiveness. %They have designed different prompt templates based on their prompting strategy and give a very straightforward table \cite[Table 1]{kim2023better}. In the analysis of their results, according to the table, one of the conclusions is that when different components ``switch on" or ``switch off", the performance of LLMs with different parameter sizes (7B and 13B) varies greatly. 
Similar findings were reported by Thomas et al.~\cite{thomas2024large} when considering prompt variations in the context of relevance labelling. Kamruzzaman\&Kim~\cite{kamruzzaman2024prompting} explored the relation between prompting strategies and social biases in LLM outputs: reduced biases were observed in more complex prompts. The most comprehensive analysis of the effect of prompt variations has been recently provided by Mizrahi et al. across a range of NLP tasks~\cite{mizrahi2024state}. These previous works, among others, have demonstrated that the structure and subtle nuances of prompt phrasing can dramatically impact the performance of the LLMs, across a wide range of tasks and contexts; however such an analysis does not exist for LLM rankers -- our work is novel as it contributes a better understanding of this phenomenon in the specific context of LLM rankers.

\begin{table*}[t]
	\caption{The PRP prompt (pointwise)~\cite{qin2023large} differs from that of RankGPT (listwise)~\cite{sun2023chatgpt} also because of accessory wordings (e.g., role-playing in RankGPT). 
		%Comparison between prompts used by PRP~\cite{qin2023large} and RankGPT~\cite{sun2023chatgpt}: they do not simply differ in the ranking approach used (PRP: pointwise; RankGPT: listwise), but also in accessory wording, e.g., role-playing in RankGPT (first line of prompt). 
		\label{prompt-wording-example}\vspace{1pt}}
	\centering \small
	\begin{tabular}{lp{300pt}}
		\toprule
		Method & Prompt                                                                                                                                                                                                                                                                                                                                                                                                                                                                                                                                                                                                                                                                                                                                   \\
		\midrule
		\makecell[cl]{PRP\\ \cite{qin2023large}} & \makecell[cl]{Passage: \{text\}\\
			Query: \{query\}\\
			Does the passage answer the query?}                                                                                                                                                                                                                                                                                                                                                                                                                                                                                                                                                                                                                                         \\ \midrule
		\makecell[cl]{RankGPT\\ \cite{sun2023chatgpt}}& \makecell[cl]{You are RankGPT, an intelligent assistant that can rank passages based on their \\relevancy to the query. I will provide you with {num} passages, each indicated by \\number identifier []. Rank the passages based on their relevance to query: \{query\}.\\
			\{PASSAGES\}\\
			Search Query: \{query\}.\\
			Rank the {num} passages above based on their relevance to the search query. The \\passages should be listed in descending order using identifiers. The most relevant \\passages should be listed first. The output format should be [] > [], e.g., [1] > [2].\\Only response the ranking results, do not say any word or explain.} \\
		\bottomrule
		\vspace{-26pt}
	\end{tabular}
\end{table*}

%\modified{Thomas et al. \cite{thomas2023large} reported that LLMs could be tailored through prompts to predict user preferences with a high degree of accuracy, demonstrating the potential for prompts to guide LLMs toward more relevant and user-aligned outputs. These findings suggest that carefully crafted prompts can bridge the gap between user queries and the information retrieved by LLMs, enhancing the relevance and utility of retrieved data. }
%\modified{Prompt design also affects LLM performance in domains such as bias mitigation and complex reasoning. Kamruzzaman and Kim \cite{kamruzzaman2024prompting}  explored how different prompting strategies could influence the presence of social biases in LLM outputs. Their research indicated that prompts designed to engage higher cognitive processes could reduce biases more effectively than simpler prompts. Additionally, Zhu et al. \cite{zhu2024inters} introduced a model that utilizes instruction tuning to refine LLM responses in search tasks, showcasing how tailored prompts can significantly enhance LLM utility in specialized applications .}
%
%\todo{conclude with importance of prompt engineering}
\vspace{-16pt}
\subsubsection{Prompt Optimisation and Self-Optimisers.}
Strategic prompt design is not only beneficial but necessary to harness the full capabilities of LLMs.
	Recent studies have taken this further by investigating LLMs as self-optimisers~\cite{yang2023large,guo2023connecting,sabbatella2024prompt}. These models utilize their generative capabilities to iteratively refine prompts, thereby enhancing their performance on downstream tasks. %This recursive enhancement process underscores the potential for LLMs to contribute actively to optimizing their interaction protocols.
	%\cite{guo2023connecting} integrated evolutionary algorithms with LLMs so that prompts can be dynamically adapted without needing gradient information. 
	%Furthermore, the integration of Bayesian Optimization for prompt tuning by Sabbatella et al. 
	%\cite{sabbatella2024prompt}'s method involves a continuous relaxation of the search space, allowing for efficient optimization of prompts; this is suitable for scenarios where only black-box access to LLMs is provided. %, enhancing the practicality of LLMs in restricted-use cases.
In our work we do not explore the adaptation of self-optimisers to prompts for LLM rankers: but our study motivates pursuing this as a future direction.

\vspace{-12pt}
\subsubsection{Zero-Shot LLM Rankers.}
We next review the four families of zero-shot generative LLM-based rankers: pointwise~\cite{zhuang2024beyond}, pairwise~\cite{qin2023large}, listwise~\cite{ma2023zero,sun2023chatgpt,tang2023found}, and setwise~\cite{zhuang2024setwise}.

%describe how each of the four types of LLM-based rankers work: pointwise~\cite{zhuang2024beyond}, pairwise~\cite{qin2023large}, listwise~\cite{ma2023zero,sun2023chatgpt,tang2023found}, and setwise~\cite{zhuang2024setwise}. 

In pointwise, two approaches can be followed:  \textit{generation} \cite{liang2022holistic,nogueira2020document} and \textit{likelihood}~\cite{zhuang2021deep,zhuang2021tilde}.  %ponte2017language,
In \textit{generation}, LLMs are prompted with a query-document pair and asked to generate a binary answer (``yes''/``no'') to the question of whether the document is relevant to the query: the likelihood of generating ``yes" (extracted from the associated token logits) decides the ranking. In \textit{likelihood}, a query likelihood model is involved in ranking. The LLM is prompted with the document and asked to produce a relevant query. Then, the ranking of documents is based on the likelihood of generating the provided query \cite{sachan2022improving}, which is obtained from the associated token logits. In our experiments with pointwise we implemented the generation approach, which is more commonly used in previous work. %as is found to be the most effective in previous work. 

Pairwise methods compare the relevance of two documents to a query, by prompting the LLM to answer which document is more relevant to the query. The ranking of documents is based on the relative preferences \cite{qin2023large}. 

For listwise ranking a list of documents and a query is passed to the LLM via the prompt. LLMs are driven to generate the document labels of relevant documents in a certain order; the ranking then depends on this \cite{ma2023zero,sun2023chatgpt,pradeep2023rankvicuna}.
%, pradeep2023rankvicuna

In setwise, a set of documents and the query are provided to the LLM, which is prompted to select the most relevant document, in an iterative way. Then sorting algorithms are used to obtain the ranking based on the preferences to documents' stopping occurs after $k$ documents have been selected (if more than $k$ were given as input) \cite{zhuang2024setwise}.

In common across all these methods is their use of LLM backbones in a zero-shot fashion, i.e. without the need for further training the LLM.

%We finally note a recent, different trend in information retrieval where generative LLMs are used to obtain dense representations of documents and queries (separately, i.e. in a bi-encoder manner), which are then in turn used for dense retrieval: while most require fine-tuning~\cite{lee2024gecko,wang2024text,wang2024improving,behnamghader2024llm2vec}, zero-shot methods are also emerging~\cite{zhuang2024promptreps}. In this paper we focus on zero-shot LLM-based re-rankers, and thus do not consider these dense retrievers; however we note they are likely affected in the same way from the issues we investigate~\cite{zhuang2024promptreps}.

\vspace{-8pt}
\section{Methodology}
\vspace{-6pt}
Different strategies for using LLMs as rankers have been proposed; they not only differ in the ranking algorithm and backbone used, but also in the specific wording provided to the LLMs to perform the task. Next, we aim to collect the original prompts used in the literature, which will serve as the foundational prompt setting for our experiments. We then analyse the individual characteristics of each prompt, building a taxonomy of the components used in the prompts along with a list of instantiations used for each component across the different prompts (Section~\ref{taxonomy}). With this information, we aim then to design experiments where we can systematically analyse the impact of components and specific wordings associated with components across different ranking algorithms and backbones. We do this by assembling prompt variations (Section~\ref{variations}). Experiment configurations are then outlined in Section~\ref{experiment_settings}.

\vspace{-8pt}
\subsection{Prompts Collection and Taxonomy of Ranking Prompt Components}
\label{taxonomy}

We started by collecting the original prompts used in current zero-shot LLM-based rankers~\cite{zhuang2024beyond,qin2023large,ma2023zero,sun2023chatgpt,tang2023found,zhuang2024setwise}. After the original prompts were collected, we analysed the prompts to identify high level components that are present in at least one original prompt, along with the associated variants. We also augmented the list of components and variants with other wordings we devised to explore specific categories further, e.g., tone words (see below). The analysis revealed five components:

\begin{table*}[t]
	\scriptsize
	\centering
	\caption{Type and wording alternatives of prompt components. None (Wording Alternative 0) refers to a prompt that does not use that component. A dash ($-$) means all prompts require that component and so the Wording Alternative 0 cannot be used (i.e. that component cannot be left empty). A checkmark means the prompt can be created without that component. }
	\label{table:classification}
	\resizebox{1\textwidth}{!}{  
		\begin{tabular}{p{35pt}|p{25pt}|p{13pt}|p{80pt}|p{80pt}|p{80pt}|p{80pt}|p{80pt}}
			\multicolumn{2}{c}{}&\multicolumn{6}{c}{Wording Alternatives} \\ \cmidrule(l){2-8}
			Component                  & Ranker    & None (0)    & 1                                                                                                                                                                                                                                                                                 & 2                                                                                                                                                  & 3                                                                      & 4                                                       & 5    \\ \midrule
			\multirow{4}{35pt}{Task Instruction (TI)} & pointwise &   -   & Does the passage answer the query?                                                                                                                                                                                                                                                & Is this passage relevant to the query?                                                                                                             & For the following query and document, judge whether they are relevant. & Judge the relevance between the query and the document. & -   \\ \cmidrule(l){2-8} 
			& pairwise  &  -    & Given a query, which of the following two passages is more relevant to the query?      &	\multicolumn{4}{c}{-}    \\ \cmidrule(l){2-8} 
			& listwise  &   -   & Rank the \{num\} passages based on their relevance to the search query.                                                                                                                                                                                                           & Sort the Passages by their relevance to the Query.   & I will provide you with \{num\} passages, each indicated by number identifier {[}{]}. Rank the passages based on their relevance to query. & \multicolumn{2}{c}{-}    \\ \cmidrule(l){2-8} 
			& setwise   &  -    & Which one is the most relevant to the query.               &\multicolumn{4}{c}{-} \\ \midrule
			\multirow{4}{35pt}{Output Type (OT)} & pointwise &      & Judge whether they are "Highly Relevant", "Somewhat Relevant", or "Not Relevant”.                                                                                                                                                                                                 & From a scale of 0 to 4, judge the relevance.                                                                                                       & Answer 'Yes' or 'No’.                                                  & Answer True/False.                                      & - \\ \cmidrule(l){2-8} 
			& pairwise  &   -   & Output Passage A or Passage B.     &    \multicolumn{4}{c}{-}\\ \cmidrule(l){2-8} 
			& listwise  &  -    &Sorted Passages = {[}   &The passages should be listed in descending order using identifiers. The most relevant passages should be listed first. The output format should be {[}{]} \textgreater {[}{]}, e.g., {[}1{]} \textgreater {[}2{]}.   &   \multicolumn{3}{c}{-} \\ \cmidrule(l){2-8} 
			& setwise   &   -   & Output the passage label of the most relevant passage.                                                                                                                                                                                                                                           & Generate the passage label.                                                                                                                      &Generate the passage label that is the most relevant to the query, then explain why you think this passage is the most relevant. & \multicolumn{2}{c}{-}                                                                                \\ \midrule
			Tone Words (TW)         &    All       & \checkmark & You better get this right or you will be punished.                                                                                                                                                                                                                                & Only output the ranking results, do not say any word or explanation.                                                                                 & Please                                                                 & Only                                                    & Must \\ \midrule
			Role Playing (RP)                      &    All       & \checkmark & You are RankGPT, an intelligent assistant that can rank passages based on their relevancy to the query.                                                                                                                                                                                                                                                                                                                            &   \multicolumn{4}{c}{-}     \\ \bottomrule
		\end{tabular}
	} \vspace{-10pt}
\end{table*}

\begin{itemize}[leftmargin=10pt, itemsep=0pt]
	\label{list:components-classification}
	\item Evidence (EV): these are the query and the associated passages to rank. 
	\item Task Instruction (TI): the instructions associated to the specific ranking strategy: these outline to the LLM the algorithmic steps to follow to produce a ranking. Example wordings include ``which passage is more relevant'' (pairwise) and ``is this passage relevant to the query'' (pointwise). 
	\item Output Type (OT): the instructions that specify the format of the output the LLM needs to generate. For example, for pointwise ranking the LLM could be instructed to generated a Yes/No or a True/False answer.
	\item Tone Words (TW): words that express a positive, negative, or neutral connotation and that help express the attitude of the prompt author towards the ranking instruction, e.g., ``please'' or ``you better get this right or you will be punished''.
	\item Role Playing (RP): a description of  the tool implemented by the LLM, used to make the LLM ``impersonate'' that role. 
\end{itemize}

Note that TI and OT largely determine the ranker family that is implemented, while TW and RP can be generally applied to any ranker family and EV are always present. Table~\ref{table:classification} lists the options we consider in our experiments for each of these components (but EV, since it is always the same). Some options refer to variations found in an original ranking prompt formulation; we augmented these with wordings we devised to explore additional alternatives for some of the components.

In addition to these components, our analysis of existing ranking prompts identified different approaches in the ordering of some of the components within the prompts; in particular:
\begin{itemize}[leftmargin=10pt, itemsep=0pt]
	\label{list:ordering-classification}
	\item Evidence Ordering (EO): the relative ordering of the query and passage(s) provided to the LLM -- whether the query is given first, followed by the passage(s), which we label as QF, or vice versa, passage(s) followed by the query (labelled PF). 
	\item Position of Evidence (PE): instruction to specify the position of the evidence in the prompt -- at the beginning (B) or at the end of the prompt (E).
\end{itemize}

\subsection{Building Prompt Variations}
\label{variations}

To explore the role of the identified components, their interactions, and the effect of specific wordings used to instantiate each of the components, we setup a large scale experiment where prompts with unique combinations of these aspects are built.

To build prompts, we first consider the ordering options we identified, and build a prompt template for each combination: this gives rise to four prompt templates, shown in Table~\ref{table:matrix}.

\begin{table}[t]
	\centering
	\caption{Prompt templates that combine the five components with the four ordering options available. Q and P denote query text passage(s) text, respectively.}
	\label{table:matrix}
%	\resizebox{0.8\textwidth}{!}{  
		\begin{tabular}{cp{120pt}p{120pt}}
			\toprule
			EO/PE & B                    & E                      \\ 
			\midrule
			QF    & RP+ TI (Q) + P + TW+ OT & RP+ TW+ OT + TI (Q) + P \\ 
			\midrule
			PF  & RP+ P + TI (Q) + TW+ OT& RP+ TW+ OT + P + TI (Q) \\ 
			\bottomrule
		\end{tabular}
%	}
	\vspace{-10pt}
\end{table}

Then, for each template, we consider all possible instantiations. The number of variations of wordings for the task instruction (TI) and output type (OT) components differ across families of rankers -- thus giving rise to different number of prompt instantiations across each family. Overall, our experiments consider 1,248 prompt variations: 768 unique prompts for pointwise; 48 for pairwise; 288 for listwise; and 144 for setwise. 

Finally note that we did minor modifications to the original prompts from previous works to improve their consistency: for example, some prompts enclosed the query in quotes, while others appended the query after a colon; other differences included the presence of multiple line breaks. We settled on adapting a unique format for these aspects. %We report these minor differences in \todo{Appendix A}. 
From preliminary experiments, we observed that these differences in prompt formatting resulted in non statistically significant differences in effectiveness: e.g., \cite{qin2023large}'s original prompt on COVID and Llama 3-8B backbone obtained an nDCG@10 of 0.8014, while our adjusted prompt lead to an nDCG@10 of 0.7966.

\vspace{-8pt}
\subsection{Experiments Settings}
\label{experiment_settings}
\vspace{-8pt}
%In the IR process, users submit a query to rankers, who then order documents by relevance. These ranked documents are then presented to the users, which shaped our experimens.
%\\

To experiment with the LLM rankers and the prompt variations, we setup a two stage ranking pipeline in line with previous work. For the first stage, we use the BM25 implementation from Pyserini~\cite{lin2021pyserini}  to retrieve the top 100 documents for a query (k1=0.9, b=0.4). For the second stage, we use the LLM ranker to re-rank these 100 documents. %: this is a common setup in previous work. 

%We use BM25 as our initial filtering stage for efficiency and benchmarking reasons. Given the potentially millions of documents and thousands of prompts in our datasets, BM25, from the Pyserini \cite{Lin_etal_SIGIR2021_Pyserini} library, significantly speeds up the process. Additionally, its common use as a baseline retriever allows for easy performance comparison to the LLMs.

%Second, we choose several LLMs and datasets as backbones. 
As LLMs backbones, we selected instruction-tuned checkpoints from the Flan-T5 family \cite{chung2024scaling}, Mistral 7B~\cite{jiang2023mistral}, and Llama3 8B~\cite{llama3modelcard}. These are popular and highly-performant open LLMs; we excluded close-source LLMs like GPT4 in our analysis because of the high costs associated with running our experiments on commercial APIs.

For Flan-T5 we considered checkpoints of different sizes: Large (783M), XL (2.85B) and XXL (11.3B). This allowed us to explore the role of backbone size in instruction following ability and ranking. We did not consider extending the experiments to larger sizes of the other backbones, e.g., Llama3 70B, because of the high computational costs associated with doing this systematically. 

We used three datasets from the ir\_datasets \cite{macavaney:sigir2021-irds} python library: TREC Deep Learning (DL) 2019 and 2020 (43 and 48 queries respectively), and BeIR TREC COVID (50 queries)~\cite{voorhees2021trec,thakur2021beir}. We performed our analysis using nDCG@10, the primary metric across these datasets, and tested statistical significance with a paired, two-tails t-test.

%Flan-T5 models have a maximum input length of 512 tokens. Upon analysis of queries and passages lengths used in our evaluation, methods such as setwise and listwise will exceed the length limits as they have multiple passages in the prompt; listwise requires the longest input length. Hence, to make comparisons between the different experimental conditions fair,we truncated inputs for all methods and LLM backbones to fit the 512 tokens limit. Based on our calculation of the longest non-query-and-passage prompt instructions, we set the query and passage lengths at 20 and 80 words, respectively. %This ensured that listwise prompts were not truncated and stayed consistent across all rankers and backbones. 

Flan-T5 models are constrained by a maximum input length of 512 tokens. By analysing the query and passage lengths within the evaluation datasets, we found that methodologies like setwise and listwise, which incorporate multiple passages within a single prompt, would exceed this limit, with listwise requiring the longest input length. To ensure equitable comparisons across different experimental setups, we uniformly truncated the inputs for all methods and LLM backbones to adhere to the 512-token constraint. Our determination of truncation lengths was based on a comprehensive analysis of the datasets employed. We calculated the length of the longest non-query-and-passage prompt instruction (pure instruction length) alongside the distribution of query and document lengths. Consequently, we standardized the query length to 20 words and the document length to 80 words. This decision was driven by the need to preserve the integrity of instructional content within the prompts, ensuring that the loss of such information would not compromise experimental consistency across different LLM backbones. In our datasets, while queries do not exceed 20 words, documents can reach up to approximately 120 words, with the majority being about 80 words. Although this truncation might result in some loss of document information, our focus was not on evaluating the absolute ranking performance but on ensuring equitable conditions for each ranker across different backbones. This strategy ensures that, even with the most extensive prompt variations and the shortest model input capabilities, no instructional wording is omitted.

\section{Results Analysis}

We base our empirical analysis along six main lines of enquiry that help us investigate the impact prompts have on LLM rankers, including what makes ranking prompts effective, how rankers respond to different prompts, how ranking methods truly compare at the net of variations in prompt wording, and the impact of LLM backbones.

\subsection{Are there better prompts?}
\begin{figure*}[p!]
	\centering
	\caption{Effectiveness (nDCG@10) of zero-shot LLM-based rankers across ranking methods, prompt variations, LLM backbones and datasets. Effectiveness achieved by original prompts for each method is marked with a star, and annotated with value.}
	\label{fig:stability}
	\begin{subfigure}{.32\linewidth}  % Adjusted from .32\textwidth to .3\columnwidth
		\centering
		\caption{DL 19}
		\includegraphics[width=\linewidth]{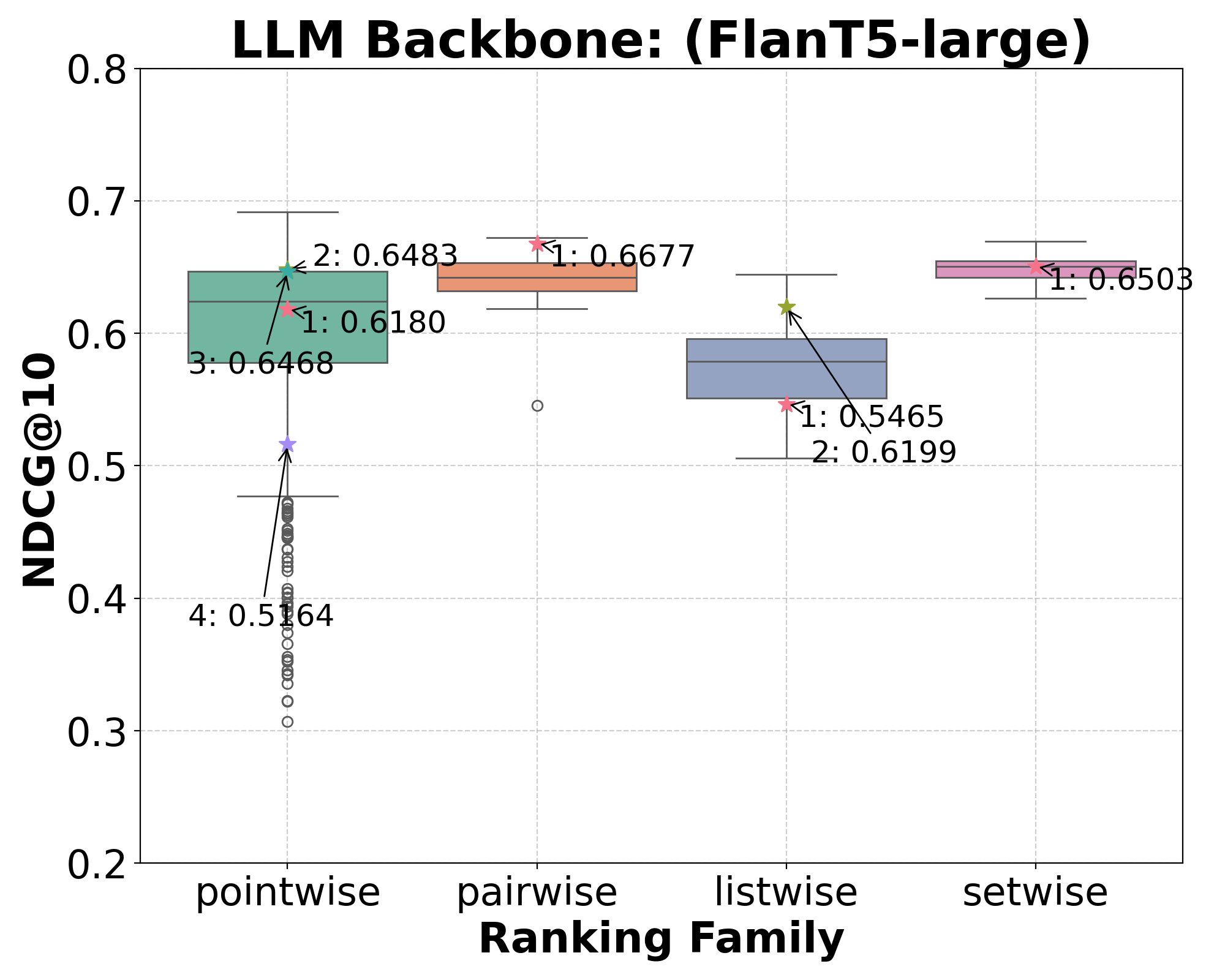}
		\label{figure:stability-sub-large-19}
	\end{subfigure}%
	\hfill
	\begin{subfigure}{.32\linewidth}  % Adjusted for fitting within a single column
		\centering
		\caption{DL 20}
		\includegraphics[width=\linewidth]{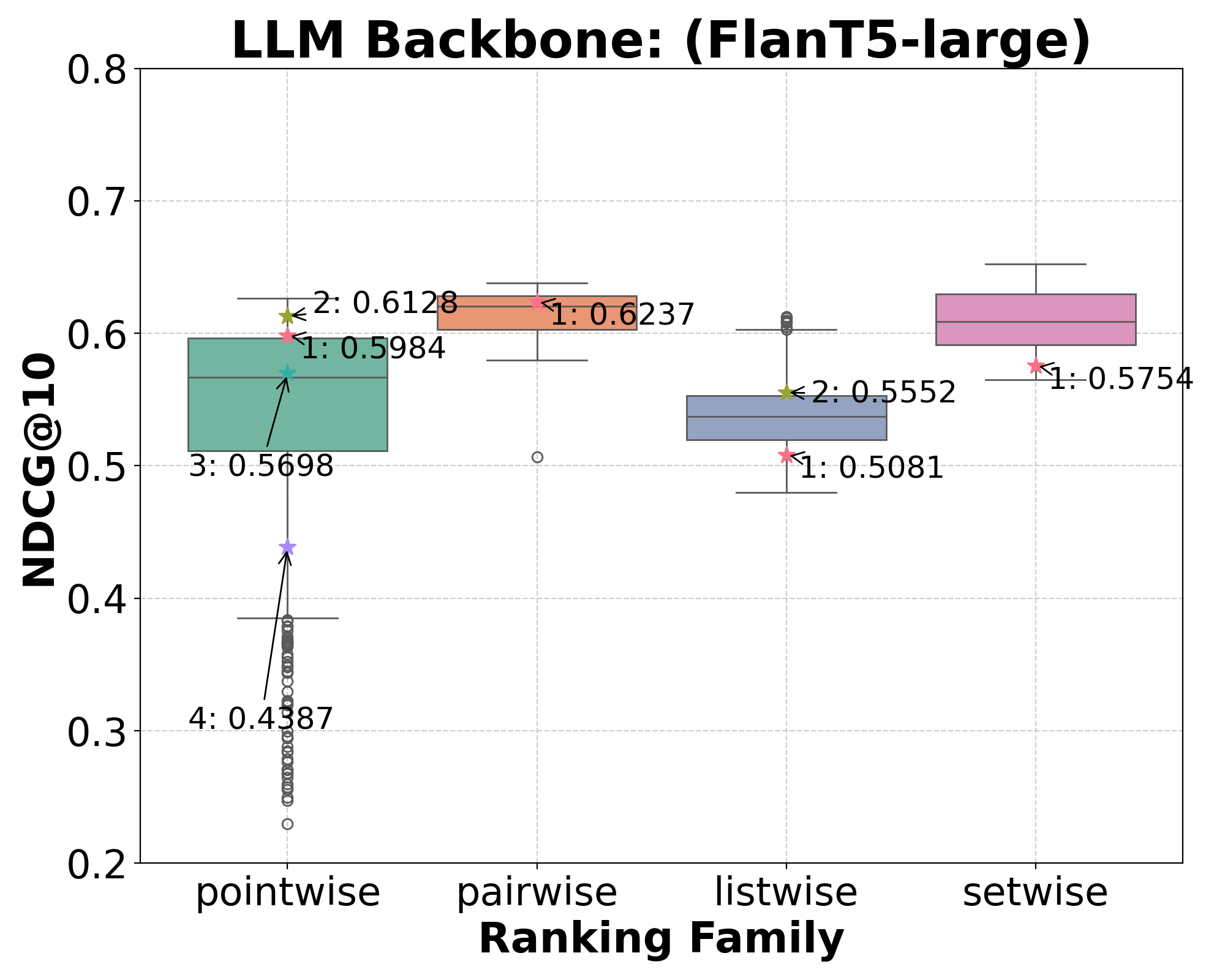}
		\label{figure:stability-sub-large-20}
	\end{subfigure}%
	\hfill
	\begin{subfigure}{.32\linewidth}  % Ensure all subfigures are in one line
		\centering
		\caption{COVID}
		\includegraphics[width=\linewidth]{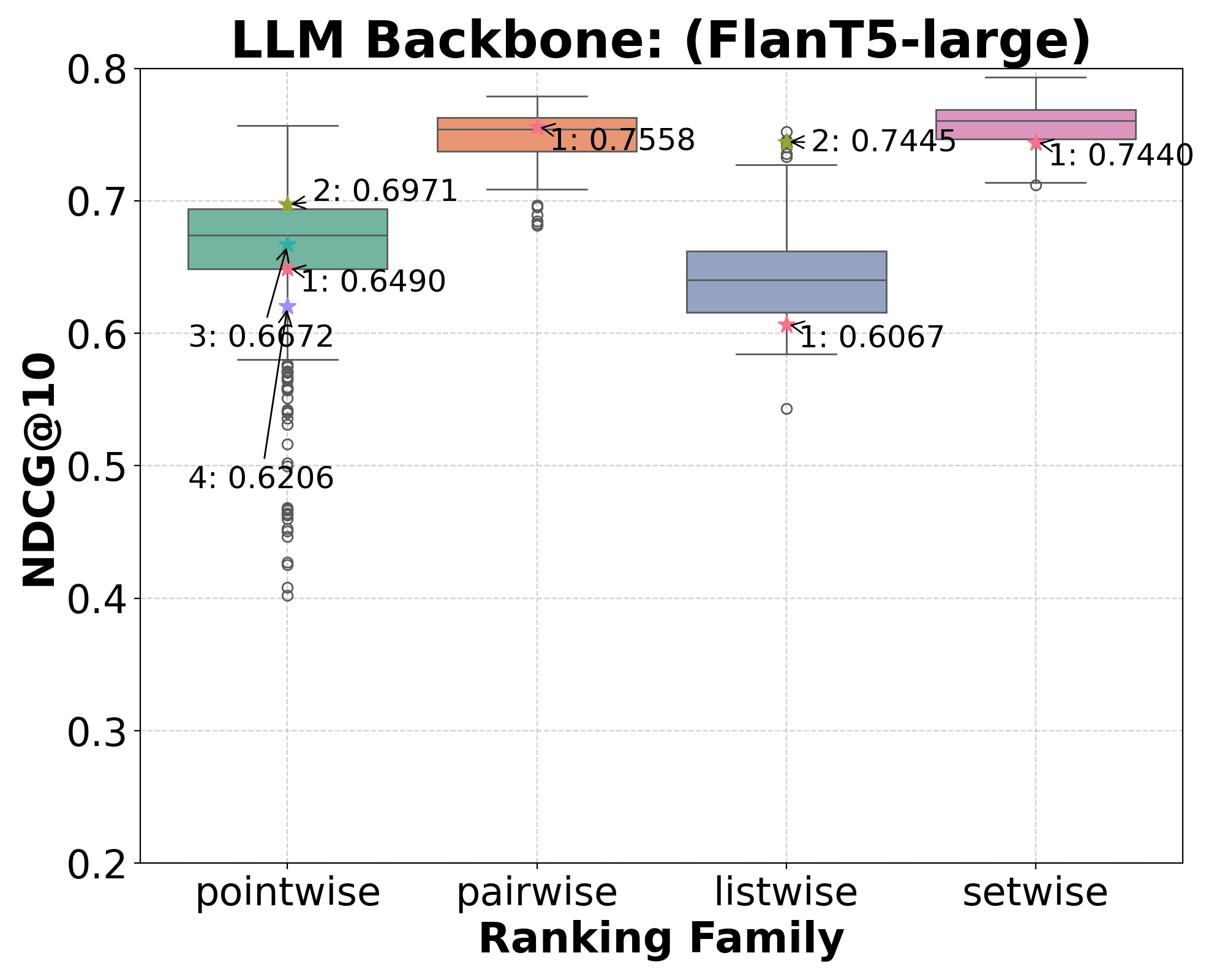}
		\label{figure:stability-sub-large-covid}
	\end{subfigure}
	
	\begin{subfigure}{.32\linewidth}  % Adjusted from .32\textwidth to .3\columnwidth
		\centering
		\includegraphics[width=\linewidth]{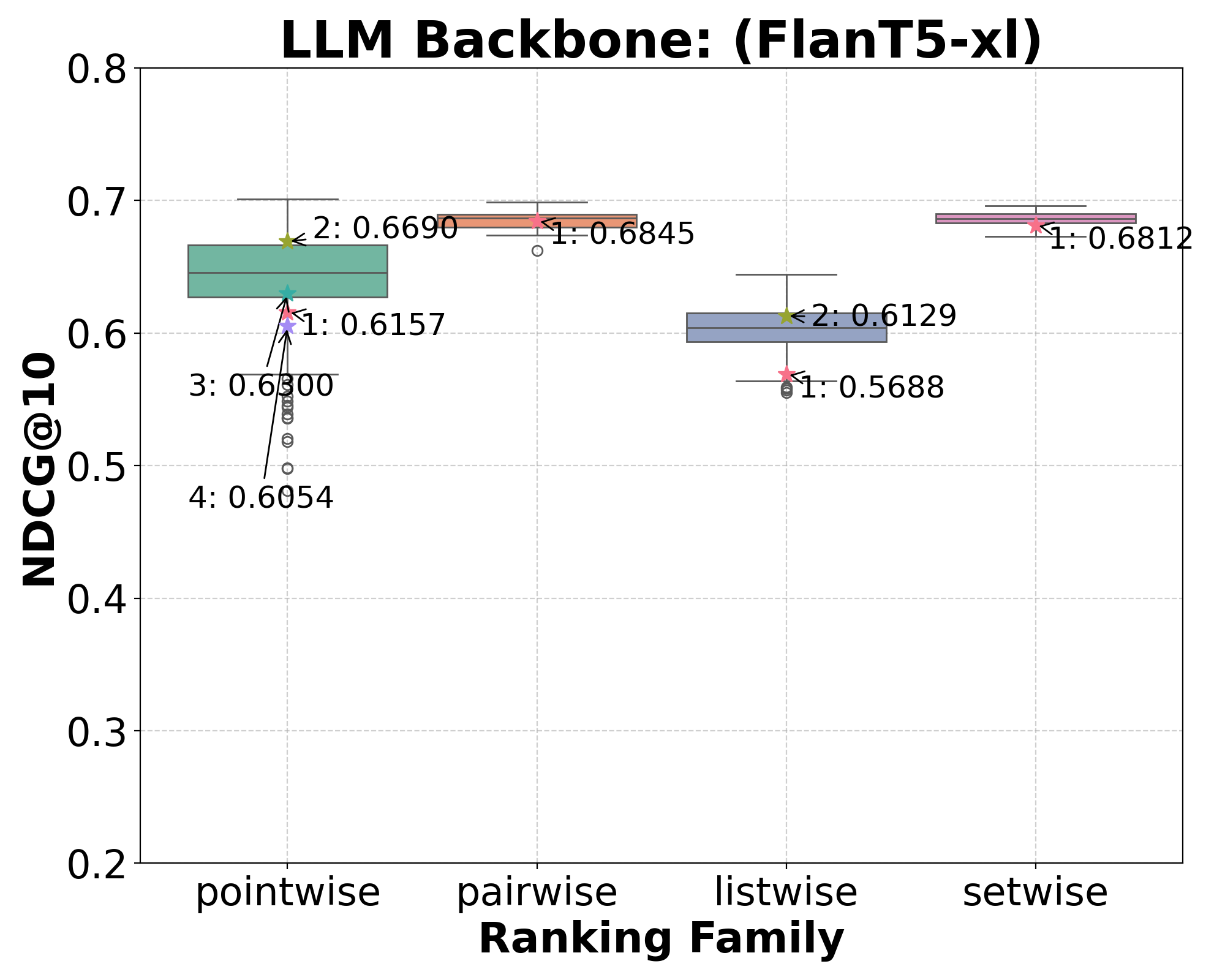}
		\label{figure:stability-sub-xl-19}
	\end{subfigure}%
	\hfill
	\begin{subfigure}{.32\linewidth}  % Adjusted for fitting within a single column
		\centering
		\includegraphics[width=\linewidth]{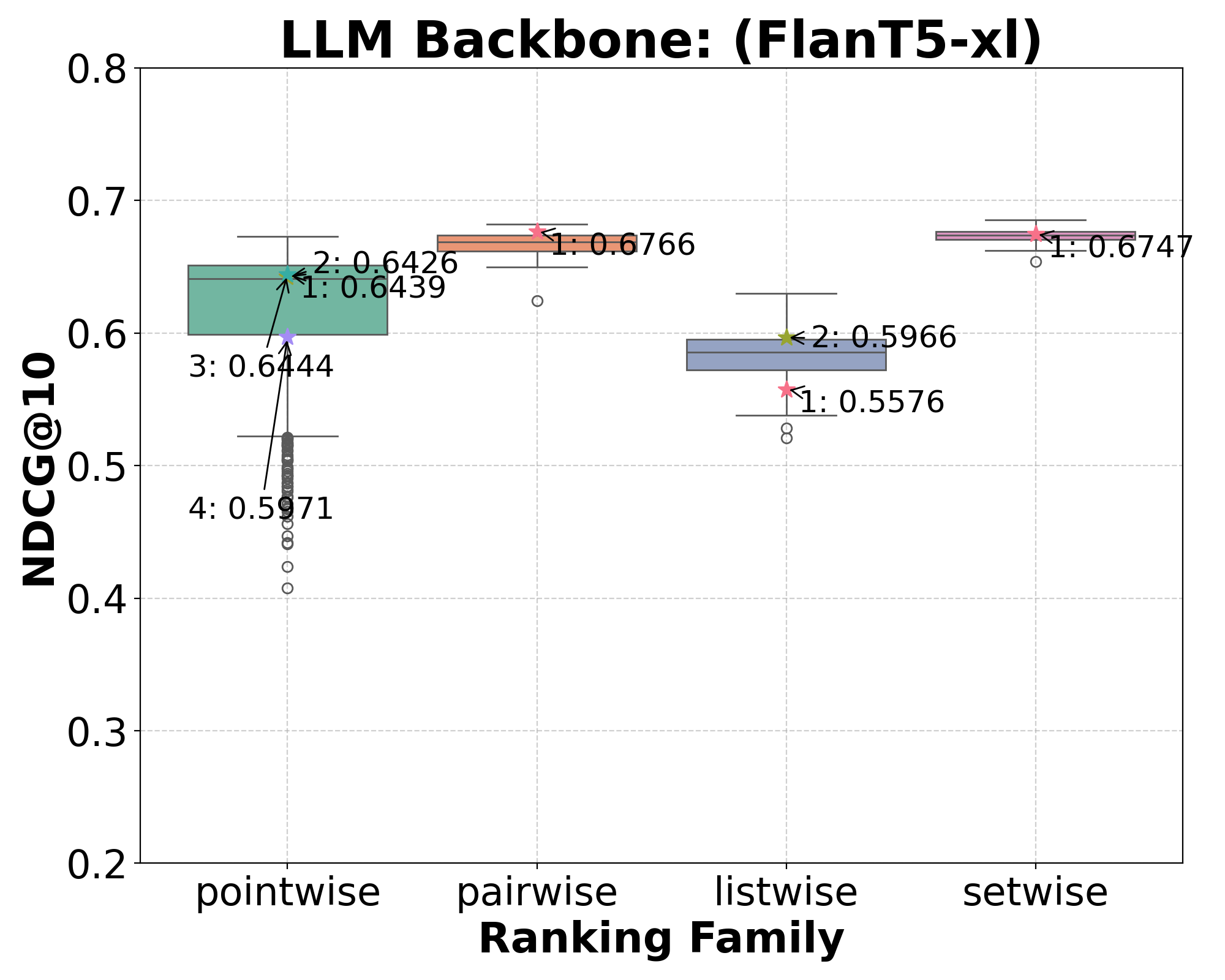}
		\label{figure:stability-sub-xl-20}
	\end{subfigure}%
	\hfill
	\begin{subfigure}{.32\linewidth}  % Ensure all subfigures are in one line
		\centering
		\includegraphics[width=\linewidth]{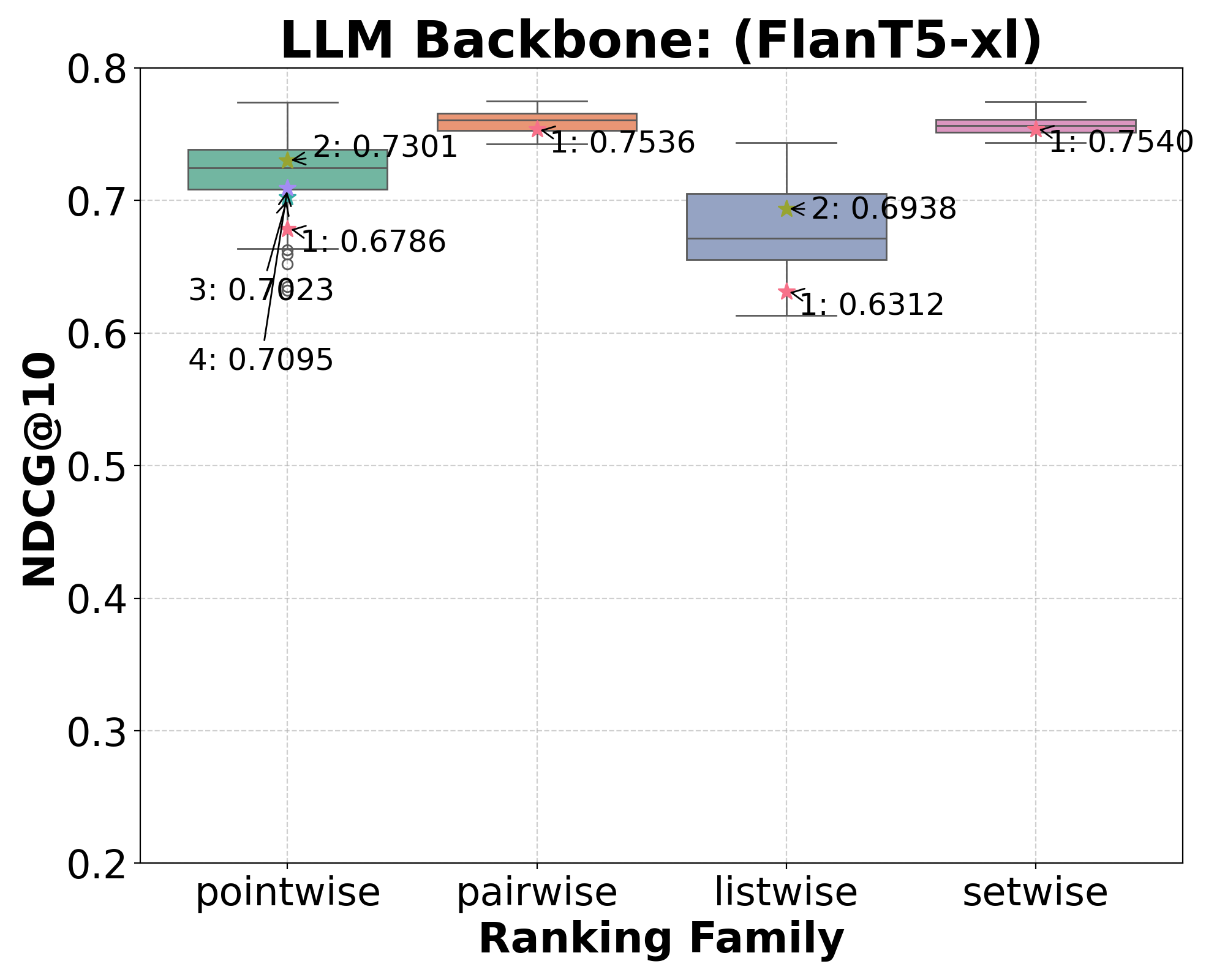}
		\label{figure:stability-sub-xl-covid}
	\end{subfigure}
	
	\begin{subfigure}{.32\linewidth}  % Adjusted from .32\textwidth to .3\columnwidth
		\centering
		\includegraphics[width=\linewidth]{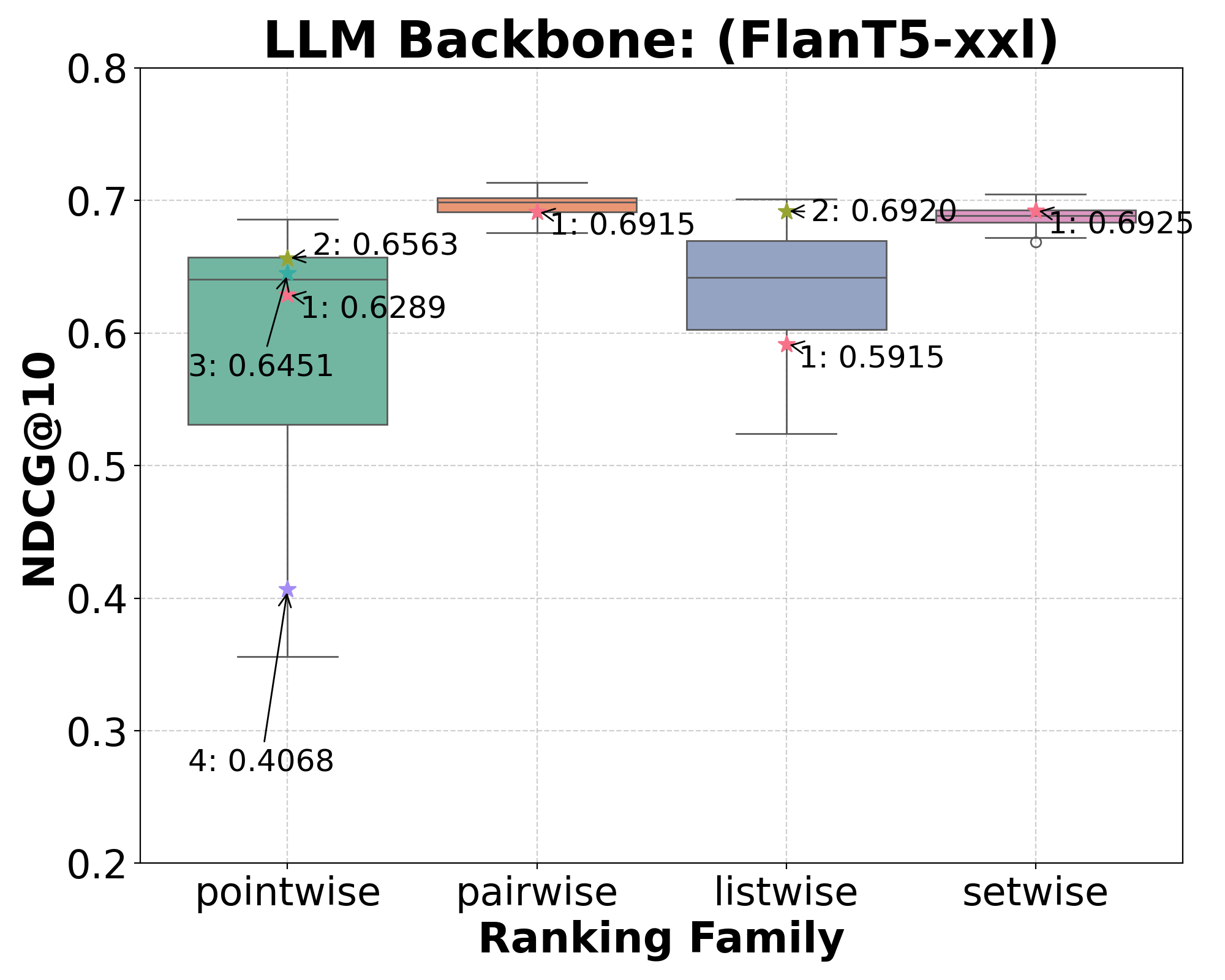}
		\label{figure:stability-sub-xxl-19}
	\end{subfigure}%
	\hfill
	\begin{subfigure}{.32\linewidth}  % Adjusted for fitting within a single column
		\centering
		\includegraphics[width=\linewidth]{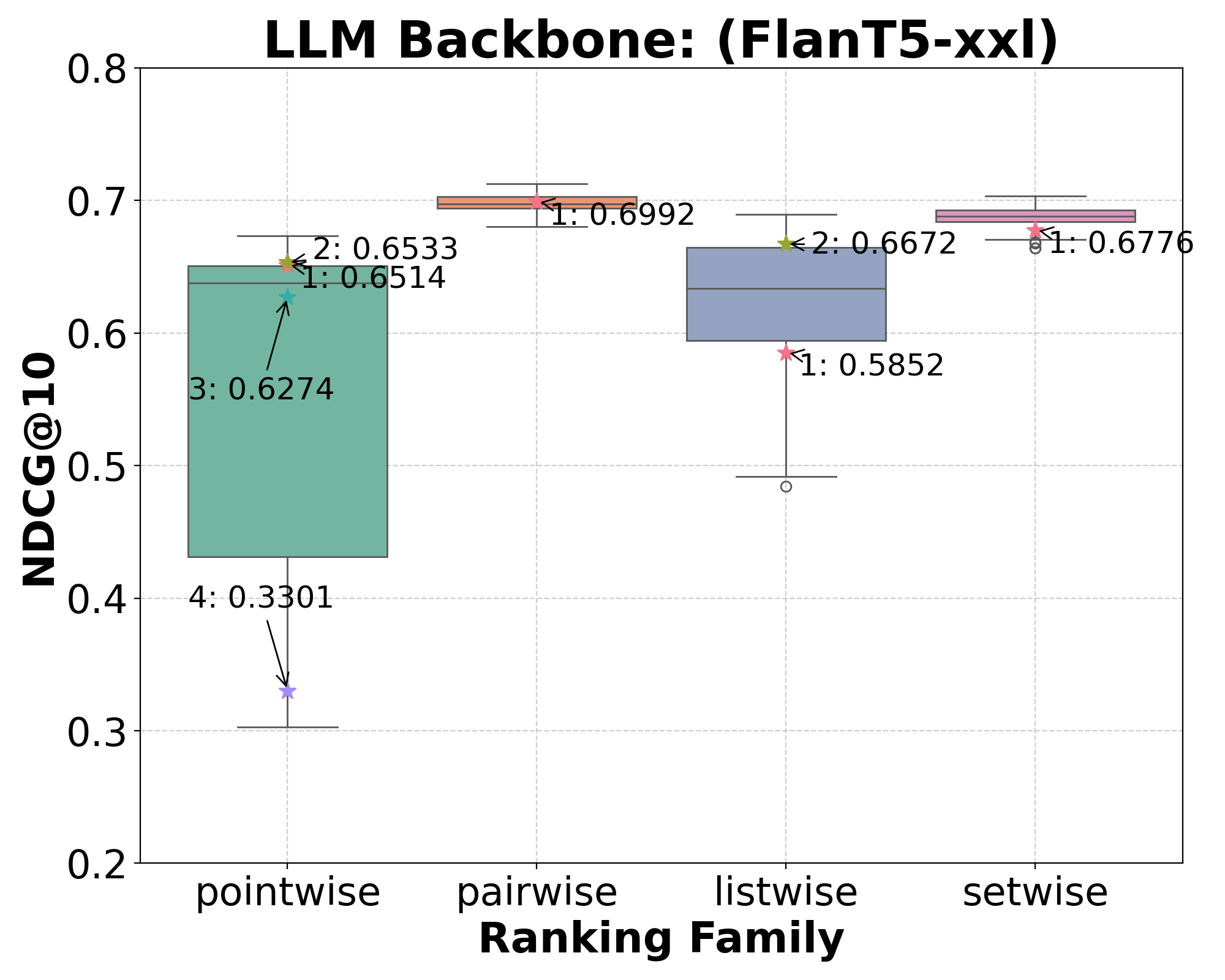}
		\label{figure:stability-sub-xxl-20}
	\end{subfigure}%
	\hfill
	\begin{subfigure}{.32\linewidth}  % Ensure all subfigures are in one line
		\centering
		\includegraphics[width=\linewidth]{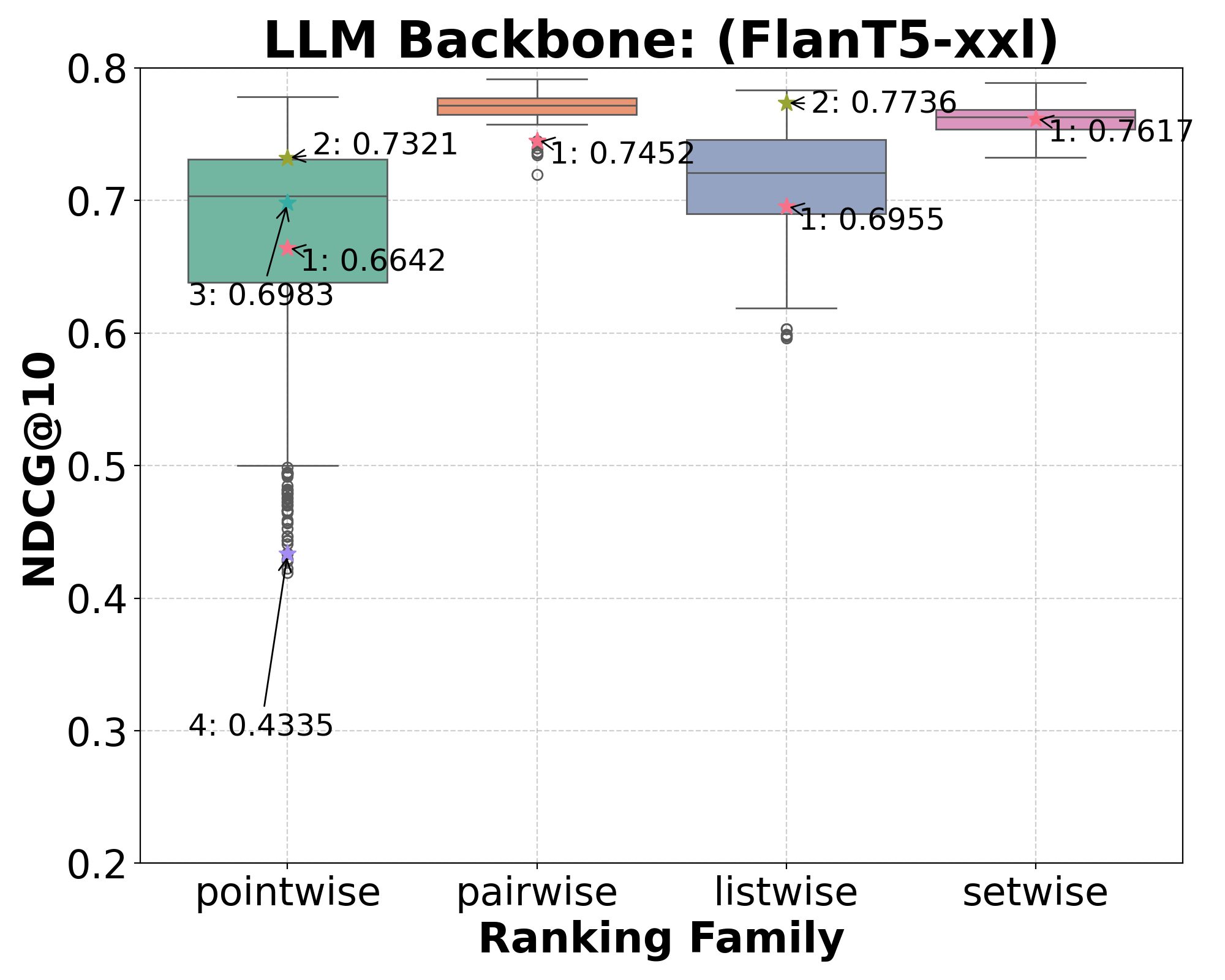}
		\label{figure:stability-sub-xxl-covid}
	\end{subfigure}
	
	\begin{subfigure}{.32\linewidth}  % Adjusted from .32\textwidth to .3\columnwidth
		\centering
		\includegraphics[width=\linewidth]{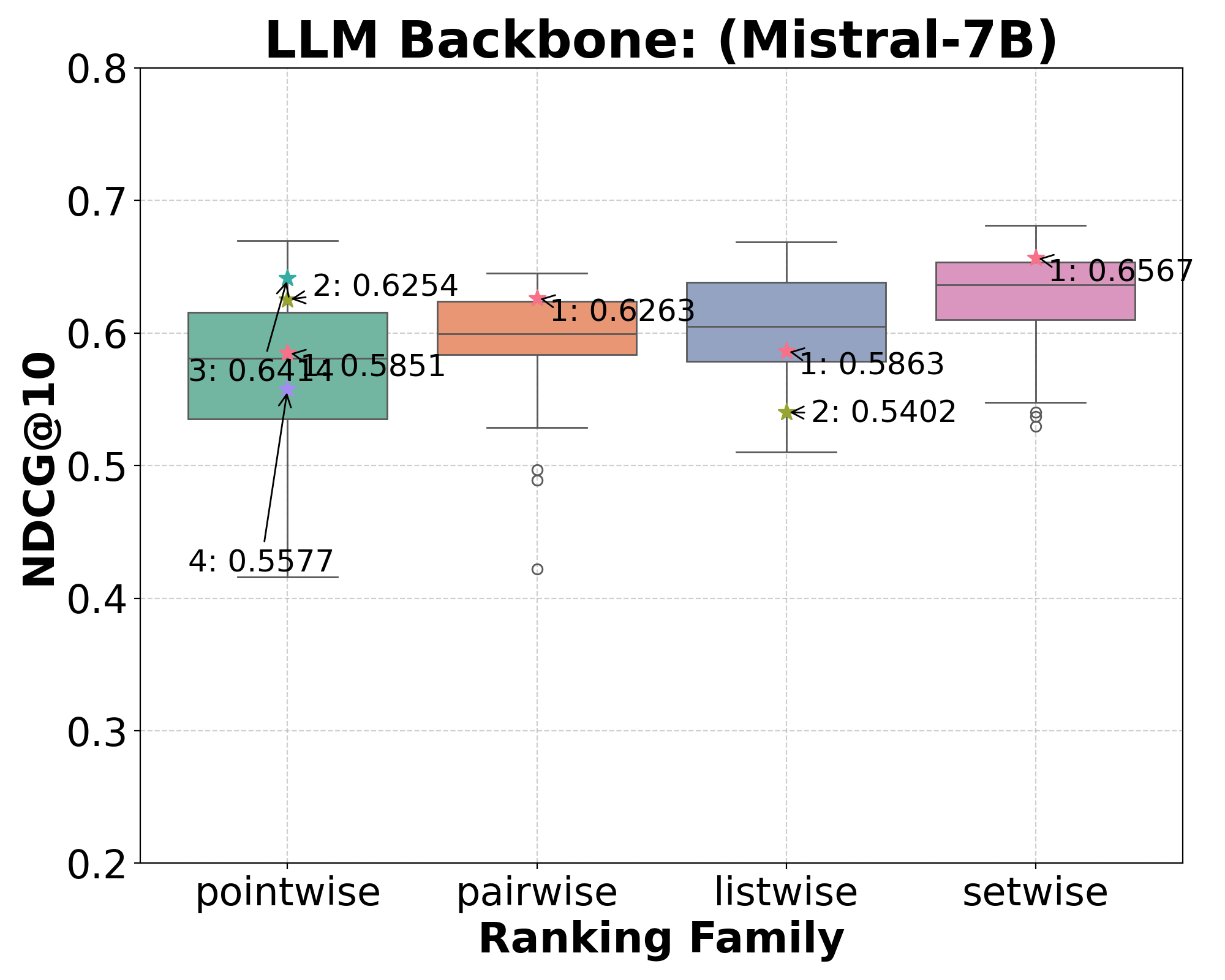}
		\label{figure:stability-sub-Mistral-7B-dl19}
	\end{subfigure}%
	\hfill
	\begin{subfigure}{.32\linewidth}  % Adjusted for fitting within a single column
		\centering
		\includegraphics[width=\linewidth]{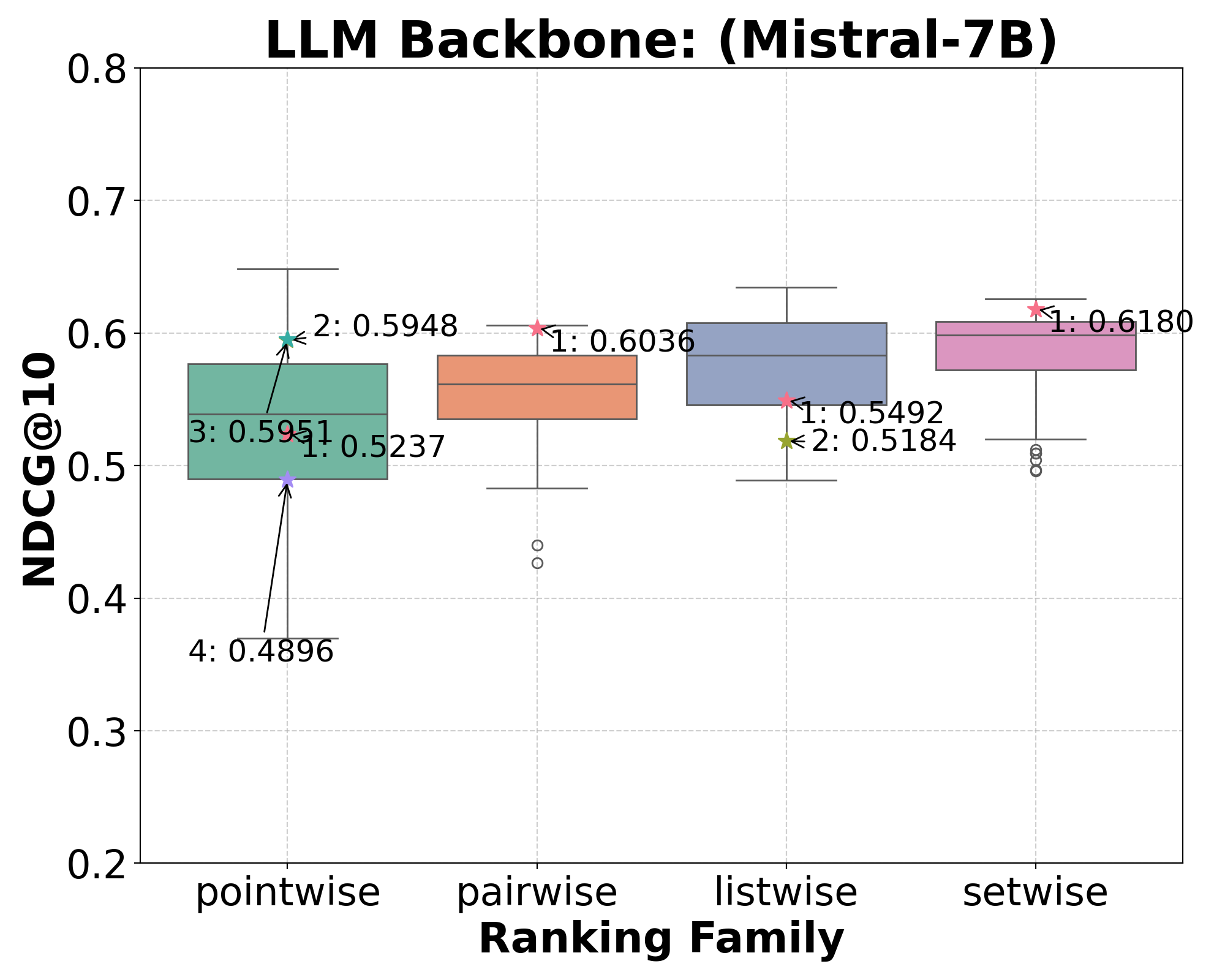}
		\label{figure:stability-sub-Mistral-7B-dl20}
	\end{subfigure}%
	\hfill
	\begin{subfigure}{.32\linewidth}  % Ensure all subfigures are in one line
		\centering
		\includegraphics[width=\linewidth]{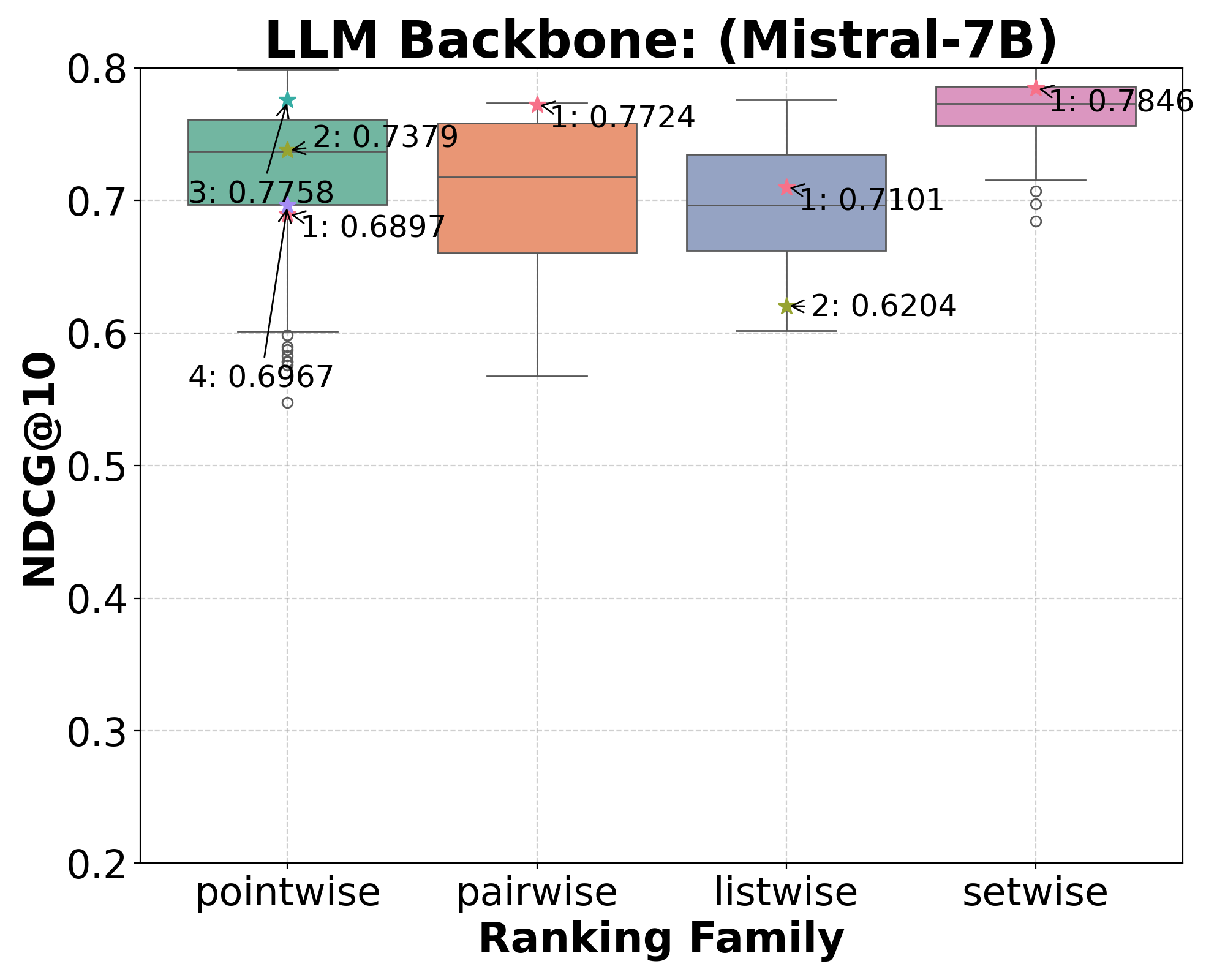}
		\label{figure:stability-sub-Mistral-7B-covid}
	\end{subfigure}
	
	\begin{subfigure}{.32\linewidth}  % Ensure all subfigures are in one line
		\centering
		\includegraphics[width=\linewidth]{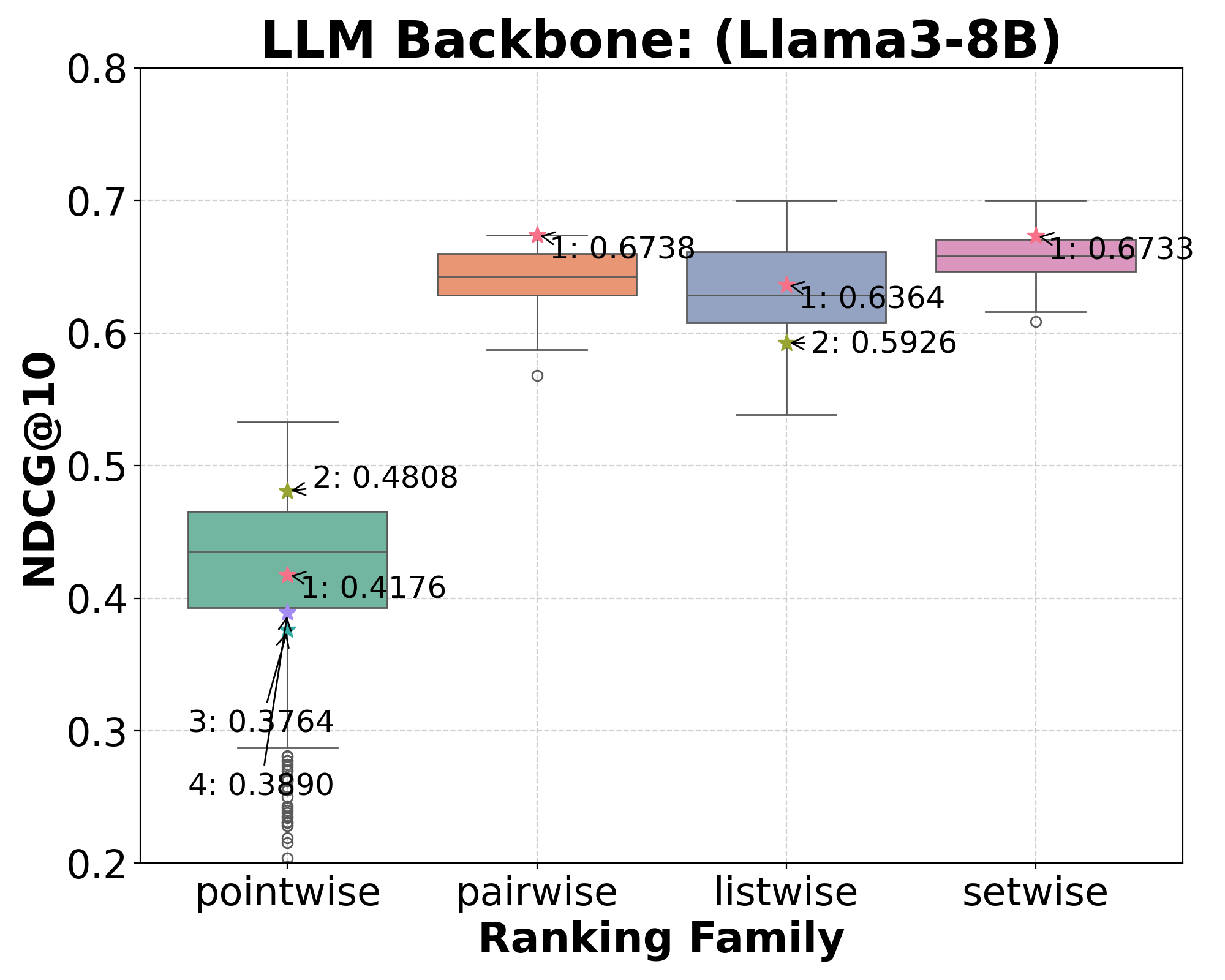}
		\label{figure:stability-sub-Llama3-8B-dl19}
	\end{subfigure}
	\begin{subfigure}{.32\linewidth}  % Ensure all subfigures are in one line
		\centering
		\includegraphics[width=\linewidth]{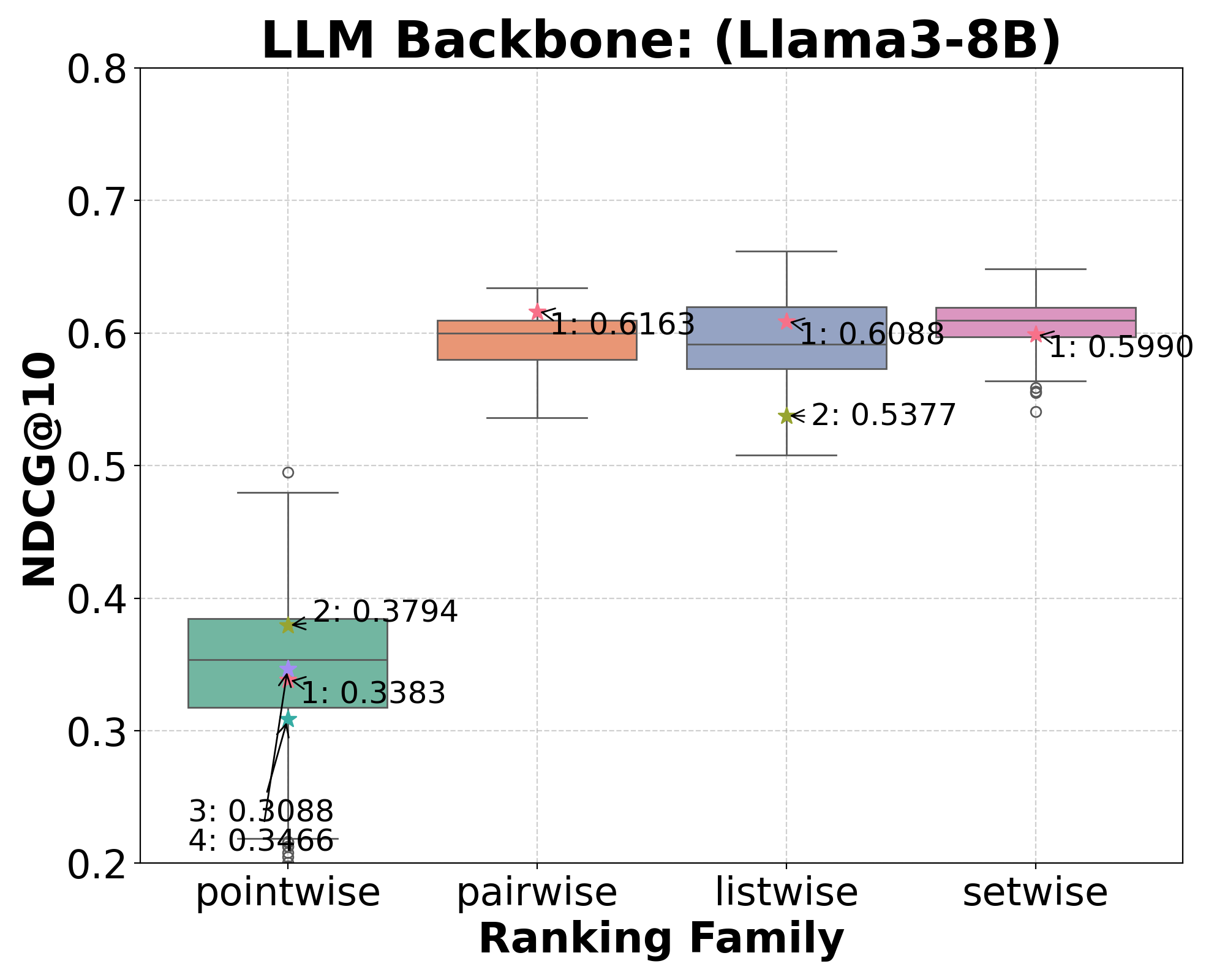}
		\label{figure:stability-sub-Llama3-8B-dl20}
	\end{subfigure}
	\begin{subfigure}{.32\linewidth}  % Ensure all subfigures are in one line
		\centering
		\includegraphics[width=\linewidth]{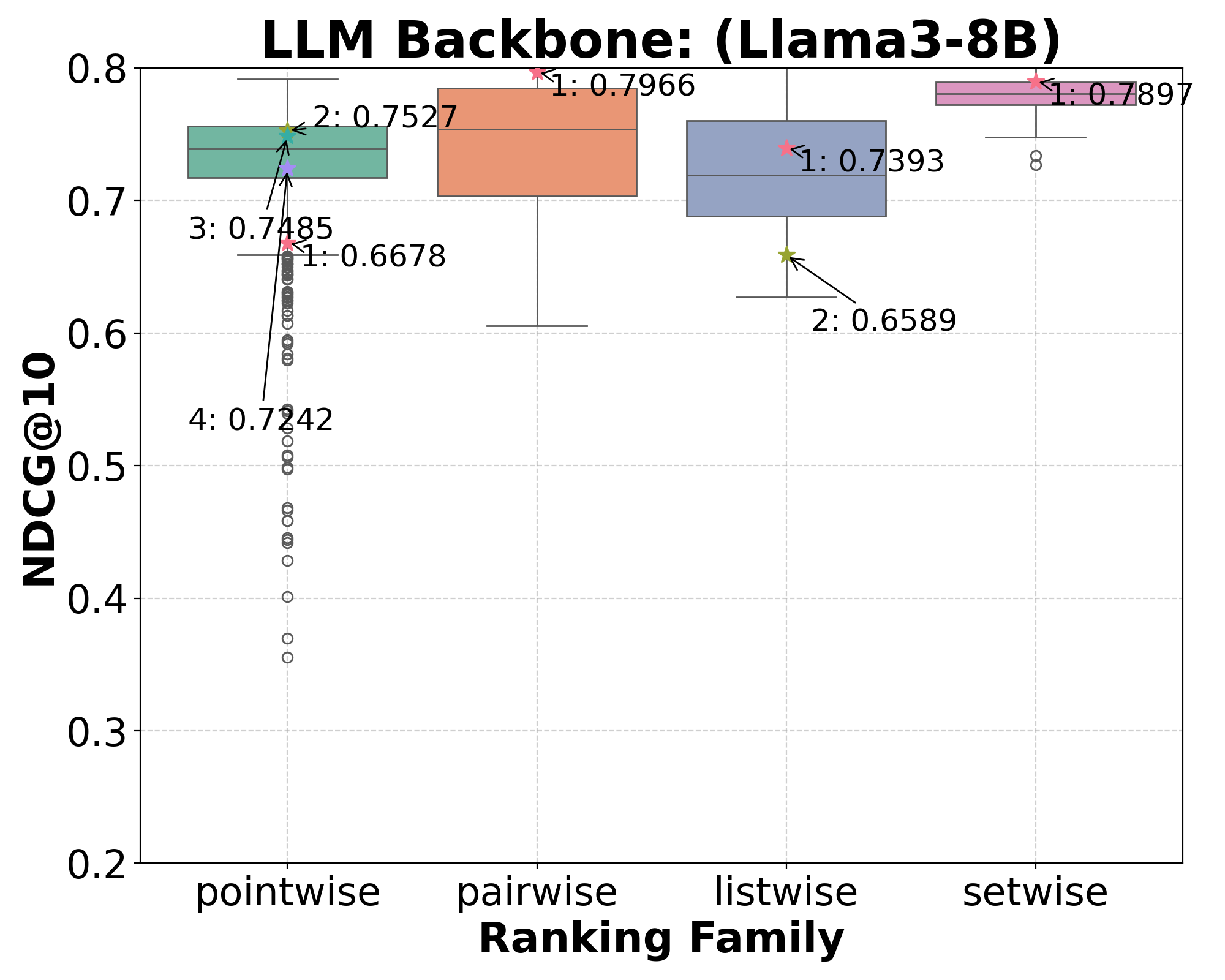}
		\label{figure:stability-sub-Llama3-8B-covid}
	\end{subfigure}
	
\end{figure*}

\vspace{-8pt}
For each family of rankers, we compare the effectiveness of the original prompts with all other prompt variations. Results are displayed in Figure~\ref{fig:stability}: for each family of rankers, the star symbols identify the original prompts, while the boxplot shows the distribution of effectiveness across all prompt variations. We found prompts that can achieve higher effectiveness than the original prompts across all cases (with most differences being statistically significant), with the exception of listwise and pairwise on COVID when the Llama 3 backbone is used. For the pairwise approach, only prompts for the FlanT5-XXL model on the COVID dataset show significant differences, likely due to the lower effectiveness variation inherent in the pairwise method. We also identify cases where the original prompt was the worst among those considered for the specific ranking family: for example for the listwise prompt evaluated on DL19 and DL 20 with the FlanT5-Large backbone. We present the actual nDCG@10 scores of original prompts and the best prompts in Table~\ref{table:original-best}.

\begin{table*}[t]
	\caption{nDCG@10 across rankers, LLMs and datasets. \textit{Po}: pointwise, \textit{Pa}: pairwise, \textit{Li}: listwise, \textit{Se}: setwise. \textit{Original}: the original (adapted) prompt. \textit{Best}: the best prompt in our experiments. For each dataset/backbone -- italics: the best performing method that used the original prompt; bold: the best performing method overall. Statistical significance differences ($^*$: $p<0.05$, $^\dagger$: $p<0.01$) between the original prompt with the best prompt measured with two-tails paired t-test.}
	\label{table:original-best}
	\centering
	\begin{subtable}{\textwidth}
		\centering
		%		\caption{Mistral-7B and Llama 3-8B}
		\resizebox{\textwidth}{!}{
			\begin{tabular}{llp{35pt}p{35pt}p{35pt}p{35pt}p{35pt}p{35pt}p{35pt}p{35pt}p{35pt}p{35pt}p{35pt}p{35pt}}
				%\toprule
				\multirow{2}{*}{} & \multirow{2}{*}{} & \multicolumn{6}{c}{Mistral-7B} & \multicolumn{6}{c}{Llama 3-8B} \\ \cmidrule{3-8} \cmidrule{9-14}
				&& \multicolumn{2}{c}{DL19} & \multicolumn{2}{c}{DL20} & \multicolumn{2}{c}{COVID} & \multicolumn{2}{c}{DL19} & \multicolumn{2}{c}{DL20} & \multicolumn{2}{c}{COVID} \\ \cmidrule{3-14}
				\multicolumn{2}{c}{Ranker}& Original & Best & Original & Best & Original & Best & Original & Best & Original & Best & Original & Best \\ \midrule
				Po & \cite{qin2023large} & 0.5851$^\dagger$ & \multirow{4}{*}{0.6699} & 0.5237$^\dagger$ & \multirow{4}{*}{\textbf{0.6486}} & 0.6897$^\dagger$ & \multirow{4}{*}{0.7988} & 0.4176$^\dagger$ &  \multirow{4}{*}{0.5328} & 0.3383$^\dagger$ &  \multirow{4}{*}{0.4950} & 0.6678$^\dagger$ &  \multirow{4}{*}{0.7916}\\ %\cmidrule{3-14}
				Po & \cite{ma2023zero} & 0.6254$^\dagger$ &  & 0.5948$^\dagger$ & & 0.7379$^\dagger$ & & 0.4808 & & 0.3794$^\dagger$ & & 0.7527 & \\ %\cmidrule{3-14}
				Po & \cite{zhuang2024beyond}.1 & \textit{0.6414} &  & 0.5951$^*$ & & 0.7758 & & 0.3764$^\dagger$ & & 0.3088$^\dagger$ & & 0.7485$^\dagger$ & \\ %\cmidrule{3-14}
				Po & \cite{zhuang2024beyond}.2 & 0.5577$^\dagger$ &  & 0.4896$^\dagger$ & & 0.6967$^\dagger$ & & 0.3890$^\dagger$ & & 0.3466$^\dagger$ & & 0.7242$^\dagger$ & \\ \midrule
				Pa & \cite{qin2023large} & 0.6263 & 0.6450 & \textit{0.6036} & 0.6059 & 0.7724 & 0.7738 & \textit{0.6738} & 0.6738 & \textit{0.6163} & 0.6340 & \textit{0.7966} & 0.8012\\ \midrule
				Li & \cite{ma2023zero} & 0.5863$^\dagger$ & \multirow{2}{*}{0.6686} & 0.5492$^\dagger$ &  \multirow{2}{*}{0.6348} & 0.7101$^\dagger$ &  \multirow{2}{*}{0.7758} & 0.6364$^\dagger$ &  \multirow{2}{*}{\textbf{0.7004}} & 0.6088$^\dagger$ &  \multirow{2}{*}{\textbf{0.6619}} & 0.7393$^\dagger$ &  \multirow{2}{*}{\textbf{0.8123}}\\ %\cmidrule{3-14}
				Li & \cite{sun2023chatgpt} & 0.5402$^\dagger$ & & 0.5184$^\dagger$ & & 0.6204$^\dagger$ & & 0.5926$^\dagger$ & & 0.5377$^\dagger$ & & 0.6589$^\dagger$ &\\ \midrule
%				Se & \cite{zhuang2023setwise} & \textit{0.6567} & \textbf{0.6811} & \textit{0.6180} & 0.6256 & \textit{0.7846} & \textbf{0.8053} & 0.6733 & 0.7000 & 0.5990& 0.6486 & 0.7897 & 0.8035\\ \midrule
				Se & \cite{zhuang2024setwise} & 0.6408$^*$ & \textbf{0.6811} & 0.5988 & 0.6256 & \textit{0.7979} & \textbf{0.8053} & 0.6654$^*$ & 0.7000 & 0.6147$^*$& 0.6486 & 0.7933 & 0.8035\\ \midrule
				%				\bottomrule
			\end{tabular}
		}
	\end{subtable}
		
	\vspace{5px}
	
	\begin{subtable}{\textwidth}
		\centering
		%		\caption{Flan-T5 Variants}
		\resizebox{\textwidth}{!}{
			\begin{tabular}{llp{30pt}p{25pt}p{30pt}p{25pt}p{30pt}p{25pt}p{30pt}p{25pt}p{30pt}p{25pt}p{30pt}p{25pt}p{30pt}p{25pt}p{30pt}p{25pt}p{30pt}p{25pt}}
				
				\multirow{2}{*}{} & \multirow{2}{*}{} & \multicolumn{6}{c}{FlanT5-Large} & \multicolumn{6}{c}{FlanT5-XL} & \multicolumn{6}{c}{FlanT5-XXL} \\ \cmidrule{3-20}
				&& \multicolumn{2}{c}{DL19} & \multicolumn{2}{c}{DL20} & \multicolumn{2}{c}{COVID} & \multicolumn{2}{c}{DL19} & \multicolumn{2}{c}{DL20} & \multicolumn{2}{c}{COVID} & \multicolumn{2}{c}{DL19} & \multicolumn{2}{c}{DL20} & \multicolumn{2}{c}{COVID} \\ \cmidrule{3-20}
				\multicolumn{2}{c}{Ranker}& Original & Best & Original & Best & Original & Best & Original & Best & Original & Best & Original & Best & Original & Best & Original & Best & Original & Best \\ \midrule
				Po & \cite{qin2023large}& 0.6180$^\dagger$ & \multirow{4}{*}{\textbf{0.6918}} & 0.5984$^*$ & \multirow{4}{*}{0.6266} & 0.6490$^\dagger$ & \multirow{4}{*}{0.7570} & 0.6157$^\dagger$ & \multirow{4}{*}{\textbf{0.7010}} & 0.6439$^*$ &\multirow{4}{*}{0.6727} & 0.6786$^\dagger$ & \multirow{4}{*}{0.7742} & 0.6289$^\dagger$ & \multirow{4}{*}{0.6860} & 0.6514 & \multirow{4}{*}{0.6733} & 0.6642$^\dagger$ & \multirow{4}{*}{0.7784}\\ 
				Po & \cite{ma2023zero} & 0.6483$^\dagger$ & & 0.6128 & & 0.6971$^\dagger$ & & 0.6690$^*$ & & 0.6426$^*$ & & 0.7301$^\dagger$ & & 0.6563& & 0.6533$^*$ & & 0.7321$^\dagger$ & \\ 
				Po & \cite{zhuang2024beyond}.1 & 0.6468$^\dagger$ & & 0.5698$^\dagger$ & & 0.6672$^\dagger$ & & 0.6300$^\dagger$& & 0.6444$^*$ & & 0.7023$^\dagger$ &  & 0.6451$^*$ & & 0.6274$^\dagger$ & & 0.6983$^\dagger$ & \\ 
				Po & \cite{zhuang2024beyond}.2 & 0.5164$^\dagger$ & & 0.4387$^\dagger$ & & 0.6206$^\dagger$ & & 0.6054$^\dagger$ & & 0.5971$^\dagger$ & &0.7095$^\dagger$ & & 0.4068$^\dagger$ & & 0.3301$^\dagger$ & & 0.4335$^\dagger$ & \\ \midrule
				Pa & \cite{qin2023large} & \textit{0.6677} & 0.6724 & \textit{0.6237} & 0.6382 & 0.7558 & 0.7788 & 0.6845 & 0.6986 & \textit{0.6766} & 0.6823 & 0.7536 & \textbf{0.7750} & 0.6915 & \textbf{0.7135} & \textit{0.6992} & \textbf{0.7126} & 0.7452$^\dagger$& \textbf{0.7917}\\ \midrule
				Li & \cite{ma2023zero}& 0.5465$^\dagger$ & \multirow{2}{*}{0.6443} & 0.5081$^\dagger$ & \multirow{2}{*}{0.6125} & 0.6067$^\dagger$ & \multirow{2}{*}{0.7521} & 0.5688$^\dagger$ & \multirow{2}{*}{0.6441} & 0.5576$^\dagger$ & \multirow{2}{*}{0.6300} & 0.6312$^\dagger$ & \multirow{2}{*}{0.7438} & 0.5915$^\dagger$ & \multirow{2}{*}{0.7011} & 0.5852$^\dagger$ & \multirow{2}{*}{0.6897} & 0.6955$^\dagger$ & \multirow{2}{*}{0.7834}\\ 
				Li & \cite{sun2023chatgpt} & 0.6199 & & 0.5552$^\dagger$ & & 0.7445 & & 0.6129$^*$ & & 0.5966$^\dagger$ & & 0.6938$^*$ & & \textit{0.6920} & & 0.6672 & & \textit{0.7736} &\\ \midrule
%				Se & \cite{zhuang2023setwise} & 0.6503 & 0.6693 & 0.5754 & \textbf{0.6525} & 0.7440 & \textbf{0.7932} & 0.6812 & 0.6959 & 0.6747 & 0.6855 & \textit{0.7540} & 0.7745 & \textit{0.6925} & 0.7047 & 0.6776 & 0.7036 & 0.7617 & 0.7890\\ 
				Se & \cite{zhuang2024setwise} & 0.6535 & 0.6693 & 0.6003$^\dagger$ & \textbf{0.6525} & \textit{0.7618}$^\dagger$ & \textbf{0.7932} &  \textit{0.6859} & 0.6959 & 0.6724 & 0.6855 & \textit{0.7543} & 0.7745 & 0.6864$^*$ & 0.7047 & 0.6679$^\dagger$ & 0.7036 & 0.7344$^\dagger$ & 0.7890\\ 
				\midrule
				
			\end{tabular}
		}
		\vspace{-18pt}
	\end{subtable}
\end{table*}

\vspace{-8pt}
\subsection{What are the characteristic of the best prompts?}
\vspace{-4pt}
Table~\ref{tab:best-settings-of-adjusted-prompt} reports the optimal prompt components for various ranking methods and datasets. We identify notable patterns as below. 

\textbf{Task Instruction and Output Type:} We analyse task instructions (TI) and output types (OT) together due to their inherent interrelation; OT largely depends on the ranking algorithm that the TI implements.
We analyse these separately for each ranking family because instructions and output types vary across ranking families. We do not analyse pairwise here as for this method there are no alternative choices for these wordings.

\textit{Pointwise ranking} does not show a consistent optimal choice; however, Task Instruction \#3 is most prevalent, while Output Type \#2 (judging relevance on a numerical scale) is seldom optimal.

\textit{Listwise ranking} shows that the best task instruction varies by dataset and LLM backbone. Output Type \#2 appears in 80\% of the most effective configurations.

%the best choice of task instruction largely depends on dataset and LLM backbone; however, choice \#2 for output type is the most prevalent among the most effective runs -- appearing in 80\% of the cases. 

\textit{Setwise ranking} studied with a single task instruction, finds Output Type \#3 (label and explain the most relevant passage) as the most common among top-performing prompts, appearing in 53\% of cases.

%\begin{itemize}[leftmargin=10pt, itemsep=0pt]
%	
%	\item \textit{Pointwise ranking} does not show a consistent optimal choice; however, Task Instruction \#3 is most prevalent, while Output Type \#2 (judging relevance on a numerical scale) is seldom optimal.
%
%	\item \textit{Listwise ranking} shows that the best task instruction varies by dataset and LLM backbone. Output Type \#2 appears in 80\% of the most effective configurations.
%	
%	%the best choice of task instruction largely depends on dataset and LLM backbone; however, choice \#2 for output type is the most prevalent among the most effective runs -- appearing in 80\% of the cases. 
%
%	\item \textit{Setwise ranking} studied with a single task instruction, finds Output Type \#3 (label and explain the most relevant passage) as the most common among top-performing prompts, appearing in 53\% of cases.
%	
%	%only one task instruction considered in setwise ranking; we observe that choice \#3 is the most frequently found among the best performing prompts, appearing 53\% of times; while choice \#2 is the least found (13\%). Choice \#3 requested to generate both the most relevant passage label and an explanation for this choice. 
%\end{itemize}

\textbf{Tone Words:} The inclusion of tone words does not present a clear trend, with the percentage of prompts achieving a higher effectiveness with no tone words or each of the five tone words distributed as follows: 18\% (no tone words), 17\%, 18\%, 20\%, 15\%, and 12\%. This indicates a relatively uniform influence of tone words choice on prompt effectiveness. 
Furthermore, no consistent patterns were observed regarding the influence of LLM backbones or datasets on the effectiveness related to tone words. Nevertheless, including a tone word in the prompt led to increased effectiveness in 82\% of the cases, underscoring the potential benefit of tone words in enhancing prompt performance.

%The inclusion of tone words does not present a clear trend, with effectiveness distributed fairly evenly across five choices. However, prompts that included a tone word showed improved effectiveness in 82\% of instances.
	
\textbf{Role Playing:} Role playing often leads to the best effectiveness for pointwise and pairwise prompts (80\% and 66\% of the cases, respectively), it has mixed effects for setwise (is present in the best prompt about half of the times), while it is not associated with the best effectiveness for listwise (13\%). Overall, 55\% of the prompts with highest effectiveness include role playing wording. Role playing was originally only used by one ranking method, RankGPT~\cite{sun2023chatgpt}, highlighting the difficulty in comparing different ranking methods if their prompt variations are not fairly explored. 
	
\textbf{Evidence Ordering:} 
%Among the best prompts, those for pointwise are characterised by a strong association with presenting to the LLM the text of the passage before that of the query (86\% of the best prompts take this form). The results are mixed instead for pairwise (40\% of the best prompts have the passage text before the query text), listwise (40\%), and setwise (46\%). Flan-T5 backbones do tend to perform best when presented with passage text before query text (66\% of cases); but results are mixed for the other backbones. 
For pointwise ranking, presenting passage text before query text is preferred in 86\% of top-performing prompts. The preference is less clear in other ranking types: 40\% for pairwise; 40\% for listwise; 46\% for setwise where the best prompts have the passage text before the query text. Considering model backbones, Flan-T5 tends to perform best when presented with passage text before query text (66\% of cases); while results are mixed for the other backbone models. 
	
\textbf{Position of Evidence:} Among the best prompts, there tend to be an overall preference for prompts that provide the evidence at the beginning (before any other instruction): this is the case in 63\% of the best prompts, with pointwise and listwise prompts exhibiting more often this pattern (73\% and 67\% respectively). Across all datasets, most best prompts for the FlanT5-XXL backbone have evidence at the beginning. This is also the case for Llama3-8B for DL19 and COVID and for Mistral-7B for COVID.

\textbf{Summary.} The analysis of the prompt templates associated with the prompts leading to the best effectiveness across all ranking families did not highlight any specific prompt wording combination that is more conducive of best effectiveness than others across all datasets and backbones -- though we found that tone words and role playing are frequent among most of the best prompts. This analysis is further extended in Section~\ref{stability} where the stability of rankers across prompts variations is considered.

\subsection{Which ranking method is most effective?}
\vspace{-6pt}
Before our analysis of prompt variations, the only comparison of ranking methods across the four families within a consistent setting of datasets and backbone was provided by Zhuang et al.~\cite{zhuang2024setwise}. According to their results, the best performing rankers were setwise and pairwise (depending on dataset and backbone), followed by listwise and then pointwise, which were distinguishably worse. 

Our analysis of prompt variations reveals a somewhat different picture. Consider the top results for each ranking family across datasets and LLM  backbone (Figure~\ref{fig:stability}). 
Similarly to previous findings, we also observe that overall pairwise and setwise methods deliver the best results. However, we find that pointwise can be as competitive as these previous methods if instructed with specific prompts. This is the case throughout all datasets and LLM backbones, with the exception of Llama 3. For Llama 3 on DL datasets, we observe that pointwise significantly underperforms other methods. We also observe that there are instances in which pointwise ranking can far exceed others: this is the case on DL19 when using FLanT5-Large and XL backbones (best pointwise nDCG@10 respectively 0.6918 and 0.7010, in most cases statistically significantly outperforming the others), and on DL20 when using Mistral-7B (best pointwise 0.6486).

%\begin{figure}[h]
%	\centering
%	\caption{Bar charts indicating the NDCG@10 scores of each ranker with the best prompts under different backbones.}
%	\label{fig:preformance-of-rankers}
%	\begin{subfigure}{.32\columnwidth}  % Adjusted from .32\textwidth to .3\columnwidth
%		\centering
%		\includegraphics[width=\linewidth]{figures/performance_of_rankers/ndcg_models_dl19.png}
%		\caption{DL19}
%	\end{subfigure}%
%	\hfill
%	\begin{subfigure}{.32\columnwidth}  % Adjusted from .32\textwidth to .3\columnwidth
%		\centering
%		\includegraphics[width=\linewidth]{figures/performance_of_rankers/ndcg_models_dl20.png}
%		\caption{DL20}
%	\end{subfigure}%
%	\hfill
%	\begin{subfigure}{.32\columnwidth}  % Adjusted from .32\textwidth to .3\columnwidth
%		\centering
%		\includegraphics[width=\linewidth]{figures/performance_of_rankers/ndcg_models_covid.png}
%		\caption{COVID}
%	\end{subfigure}%
%	
%\end{figure}

\begin{table*}[h!]
	\centering
	\vspace{-22pt}
	\caption{Prompt templates containing the components that lead to the most effective results across combinations of ranking method, LLM backbones, and dataset. The notation used corresponds to component definitions in Section \ref{taxonomy} and options in Table \ref{table:classification} (e.g., TI\_3 for \textit{Task Instruction} with the \textit{Third Option}).}
	\label{tab:best-settings-of-adjusted-prompt}
	\vspace{-6pt}
	\begin{subtable}{\textwidth}
		\centering
		\caption{Dataset: DL19}
		\resizebox{0.8\textwidth}{!}{
			\begin{tabular}{@{}lc|c|c|c@{}}
				\toprule
				Model & Pointwise & Pairwise & Listwise & Setwise \\ \midrule
				FlanT5-large & \makecell{TI\_3, OT\_1, \\ TW\_0, PF, B, RP\_1} & \makecell{TI\_1, OT\_1, \\ TW\_4, QF, B, RP\_None} & \makecell{TI\_3, OT\_2, \\ TW\_2, PF, E, RP\_None} & \makecell{TI\_1, OT\_3, \\ TW\_0, QF, B, RP\_None} \\ \midrule
				FlanT5-xl & \makecell{TI\_2, OT\_3, \\ TW\_4, PF, E, RP\_1} & \makecell{TI\_1, OT\_1, \\ TW\_2, QF, B, RP\_1} & \makecell{TI\_3, OT\_2, \\ TW\_3, QF, B, RP\_None} & \makecell{TI\_1, OT\_3, \\ TW\_3, QF, E, RP\_1} \\ \midrule
				FlanT5-xxl & \makecell{TI\_4, OT\_3, \\ TW\_1, PF, B, RP\_None} & \makecell{TI\_1, OT\_1, \\ TW\_0, QF, E, RP\_1} & \makecell{TI\_1, OT\_2, \\ TW\_2, QF, B, RP\_None} & \makecell{TI\_1, OT\_1, \\ TW\_4, QF, B, RP\_1} \\ \midrule
				Mistral-7B & \makecell{TI\_1, OT\_4, \\ TW\_3, PF, E, RP\_1} & \makecell{TI\_1, OT\_1, \\ TW\_0, QF, E, RP\_1} & \makecell{TI\_1, OT\_1, \\ TW\_2, QF, B, RP\_None} & \makecell{TI\_1, OT\_3, \\ TW\_0, QF, E, RP\_1} \\ \midrule
				Llama3-8B & \makecell{TI\_2, OT\_3, \\ TW\_2, PF, B, RP\_1} & \makecell{TI\_1, OT\_1, \\ TW\_0, QF, B, RP\_None} & \makecell{TI\_1, OT\_2, \\ TW\_0, QF, B, RP\_None} & \makecell{TI\_1, OT\_1, \\ TW\_3, PF, E, RP\_None} \\ 
				\bottomrule
			\end{tabular}
		}
	\end{subtable}
	
	%\vspace{1em} % Add some vertical space between the subtables
	
	\begin{subtable}{\textwidth}
		\centering
		\caption{Dataset: DL20}
		\resizebox{0.8\textwidth}{!}{
			\begin{tabular}{@{}lcccc@{}}
				\toprule
				Model & Pointwise & Pairwise & Listwise & Setwise \\ \midrule
				FlanT5-large & \makecell{TI\_3, OT\_3, \\ TW\_3, PF, B, RP\_1} & \makecell{TI\_1, OT\_1, \\ TW\_2, PF, E, RP\_None} & \makecell{TI\_3, OT\_2, \\ TW\_1, PF, E, RP\_None} & \makecell{TI\_1, OT\_1, \\ TW\_2, PF, E, RP\_1} \\ \midrule
				FlanT5-xl & \makecell{TI\_3, OT\_3, \\ TW\_4, QF, B, RP\_1} & \makecell{TI\_1, OT\_1, \\ TW\_5, QF, B, RP\_1} & \makecell{TI\_2, OT\_2, \\ TW\_3, PF, E, RP\_1} & \makecell{TI\_1, OT\_2, \\ TW\_5, PF, B, RP\_None} \\ \midrule
				FlanT5-xxl & \makecell{TI\_3, OT\_1, \\ TW\_3, PF, B, RP\_None} & \makecell{TI\_1, OT\_1, \\ TW\_2, PF, E, RP\_1} & \makecell{TI\_2, OT\_2, \\ TW\_3, PF, B, RP\_None} & \makecell{TI\_1, OT\_3, \\ TW\_3, PF, B, RP\_1} \\ \midrule
				Mistral-7B & \makecell{TI\_1, OT\_4, \\ TW\_2, PF, E, RP\_1} & \makecell{TI\_1, OT\_1, \\ TW\_1, QF, E, RP\_None} & \makecell{TI\_3, OT\_1, \\ TW\_0, QF, B, RP\_1} & \makecell{TI\_1, OT\_3, \\ TW\_2, QF, E, RP\_1} \\ \midrule
				Llama3-8B & \makecell{TI\_1, OT\_3, \\ TW\_1, PF, B, RP\_1} & \makecell{TI\_1, OT\_1, \\ TW\_3, PF, E, RP\_1} & \makecell{TI\_3, OT\_2, \\ TW\_0, QF, B, RP\_None} & \makecell{TI\_1, OT\_2, \\ TW\_4, PF, E, RP\_None} \\
				\bottomrule
			\end{tabular}
		}
	\end{subtable}
	
	%\vspace{1em} % Add some vertical space between the subtables
	
	\begin{subtable}{\textwidth}
		\centering
		\caption{Dataset: COVID}
		\resizebox{0.8\textwidth}{!}{
			\begin{tabular}{@{}lcccc@{}}
				\toprule
				Model & Pointwise & Pairwise & Listwise & Setwise \\ \midrule
				FlanT5-large & \makecell{TI\_3, OT\_1, \\ TW\_1, PF, B, RP\_1} & \makecell{TI\_1, OT\_1, \\ TW\_5, PF, B, RP\_1} & \makecell{TI\_3, OT\_2, \\ TW\_4, PF, E, RP\_None} & \makecell{TI\_1, OT\_1, \\ TW\_1, PF, B, RP\_None} \\ \midrule
				FlanT5-xl & \makecell{TI\_3, OT\_1, \\ TW\_1, PF, B, RP\_1} & \makecell{TI\_1, OT\_1, \\ TW\_5, PF, B, RP\_1} & \makecell{TI\_2, OT\_2, \\ TW\_1, QF, E, RP\_None} & \makecell{TI\_1, OT\_3, \\ TW\_5, QF, E, RP\_None} \\ \midrule
				FlanT5-xxl & \makecell{TI\_3, OT\_1, \\ TW\_1, PF, B, RP\_None} & \makecell{TI\_1, OT\_1, \\ TW\_0, PF, B, RP\_1} & \makecell{TI\_1, OT\_2, \\ TW\_3, PF, B, RP\_None} & \makecell{TI\_1, OT\_3, \\ TW\_5, PF, B, RP\_1} \\ \midrule
				Mistral-7B & \makecell{TI\_2, OT\_2, \\ TW\_4, PF, B, RP\_1} & \makecell{TI\_1, OT\_1, \\ TW\_3, QF, B, RP\_1} & \makecell{TI\_1, OT\_1, \\ TW\_4, QF, B, RP\_None} & \makecell{TI\_1, OT\_1, \\ TW\_2, QF, B, RP\_None} \\ \midrule
				Llama3-8B & \makecell{TI\_2, OT\_3, \\ TW\_5, QF, E, RP\_1} & \makecell{TI\_1, OT\_1, \\ TW\_4, QF, B, RP\_None} & \makecell{TI\_1, OT\_2, \\ TW\_0, QF, B, RP\_None} & \makecell{TI\_1, OT\_3, \\ TW\_1, QF, B, RP\_1} \\ 
				\bottomrule
			\end{tabular}
		}
	\end{subtable}
	\vspace{-20pt}
\end{table*}

\vspace{-4pt}
\subsection{Are ranking methods stable?}
\vspace{-6pt}
\label{stability}
We consider the four ranking families independently, and study the variance of their effectiveness across all prompt variations for that family. Results are displayed in Figure~\ref{fig:stability}. We observe that pointwise methods display the largest variability in effectiveness due to prompt variations, with some prompt variations delivering poor effectiveness. Setwise and pairwise instead do better, displaying lower variability: setwise achieves this across all backbones and datasets investigated, while pairwise displays larger variability in specific conditions, e.g., when using Mistral, and for COVID when using Llama3.

Two key insights arise: 1) ranking algorithms are susceptible to prompt variations and may exhibit wide-ranging performance variations, and 2) different ranking algorithms have varying sensitivity to prompts, indicating the potential for some to mitigate the impact of these variations. This suggests that more comprehensive prompting optimisation, e.g., via self-prompting, may further improve effectiveness across all rankers.

\vspace{-4pt}
\subsection{Does the LLM size matter?}
\vspace{-6pt}
We answer this question by focusing on the FlanT5 models only, which come in three different sizes (first three rows of plots in Figure~\ref{fig:stability}). We observe that in general larger models deliver higher effectiveness and reduced variations across the board. However, the pointwise family represents an exception. Improvements are observed when passing from FlanT5-large to FlanT5-XL. However when using FlanT5-XXL we observe both decreased effectiveness and large variance in effectiveness across the prompt variations.

%In TABLE \ref{table:original-best}, We can see some consistently better performance if the model bigger. For example, in the pairwise, COVID dataset, scores improve from Large (the original: 0.7417, the best: 0.7788) to XL (0.7581, 0.7750) to XXL (0.7625, 0.7917). However, sometimes, it becomes uncertain; for instance, the highest score of the XXL model is even lower than any of the others in the pointwise with DL19 dataset.
%
%For most scenarios and datasets analyzed, Flan-T5-XXL tends to offer the best performance, followed by Flan-T5-XL and then Flan-T5-Large. This analysis suggests that model size indeed matters, with larger models (XXL in particular) generally performing better than smaller ones (XL and Large). However, the extent of this impact varies across different datasets and ranker types.

\vspace{-4pt}
\subsection{Does the LLM backbone matter?}
\vspace{-6pt}
We compare three backbones across the results from Figure~\ref{fig:stability}): FlanT5-XXL (11.3B parameters), Mistral-7B and Llama3-8B: the last two are of comparable size, while the first has approximately 40-60\% more parameters. We observe that generally Llama3 and FlanT5 outperform Mistral-based rankers. Llama3 and FlanT5 have overall similar effectiveness, though on COVID Llama3-based rankers consistently outperform those with FlanT5. We also observe that Mistral-based rankers exhibit larger variance in effectiveness due to prompt variations than rankers based on the other backbones.

\vspace{-4pt}
\section{Further Discussion and Limitations}
\vspace{-6pt}

Latency is an important factor affecting the deployment of rankers in production. We did not consider query latency as one of the dimensions of analysis and comparison because our focus was on varying the prompts to measure ranking effectiveness. Zhuang et al.~\cite{zhuang2024setwise} provided a first insight into comparing latency among the zero-shot LLM rankers we considered. Latency of the methods and prompts we considered in our analysis would be similar to that performed there; however we note that some of the prompts we consider are longer than others (e.g., if the role playing component is added to prompts) -- longer prompts result in increased query latency.

Overall, our experiments considered 1,248 prompt variations. While this is a large number, many more prompt variations could have been designed and investigated (and the space of prompt variations is virtually infinite). However, sensibly increasing the number of prompt variations in our experiments would have been infeasible as it would have exceeded our compute budget. 
We ran these prompts across three datasets, and using five different LLMs with up to 11B parameters. The execution of these experiments required a large amount of compute power; we ran experiments across three clusters: one with Nvidia A100 GPUs, another with Nvidia H100 GPUs, and a smaller one with Nvidia A6000 GPUs.
Overall, the experiments took in excess of 12,400 GPU-hours to execute. Varying prompt for the pairwise ranking results in large compute requirements compared to other methods because of the extensive number of pairwise computations (and thus LLM inferences) required by this method to answer a query.  Although we only considered zero-shot LLM ranking approaches and thus did not conduct any LLM training, we recognize that our experiments may have still consumed substantial energy, thereby contributing to CO2 emissions~\cite{scells2022reduce} and water consumption~\cite{zuccon2023beyond}.

In our analysis we considered instruction-tuned checkpoints from three LLM backbones. %: FlanT5 (in three different sizes), Mistral and Llama 3. 
A wider range of LLMs could have been considered, importantly including OpenAI's GPT models such as GPT3.5 and GPT4. GPT3.5 for example was found more effective than Llama 2 and Vicuna based LLM rankers in a limited set of experiments in previous work~\cite{zhuang2024setwise}. We were however restricted to using non-commercial models because of the high costs involved in performing our large number of experiments with commercial APIs. For example, using the cost estimates from \cite{zhuang2024setwise}, executing our experiments for pairwise, listwise and setwise across all considered prompts using GPT3.5 alone would have costed approximately USD \$5,000.

%In our experiments we considered a large number of prompt variations: for pointwise we consider 768 unique prompts; for pairwise 48; for listwise 288; and for setwise 144 -- for a total of 1,248 prompts. We ran these prompts across three datasets, and using five different LLMs with up to 11B parameters. The execution of these experiments required a large amount of compute power; we ran experiments across three clusters: one with Nvidia A100 GPUs, another with Nvidia H100 GPUs, and a smaller one with Nvidia A6000 GPUs. Although we only considered zero-shot LLM ranking approaches and thus did not conduct any LLM training, we recognize that our experiments may have still consumed substantial energy, thereby contributing to CO2 emissions~\cite{scells2022reduce} and water consumption~\cite{zuccon2023beyond}.

%The publicly available LLMs we used may have been trained with content that contains several types of societal biases~\cite{gallegos2024bias} -- and these biases  may permeate in the rankings our zero-shot LLM rankers produced. Future research could explore ways to mitigate these biases prompt engineering.

%0.171*48*(43+48+50) + 0.045*288*(43+48+50) + 0.084*144*(43+48+50)

\vspace{-8pt}
\section{Conclusions}
\vspace{-4pt}
Recent works have shown the promise of using generative LLMs to implement effective zero-shot re-rankers. 
%While at first different methods differ because of the ranking algorithm they implement, we observe that they do also differ from some of the characteristics of the prompts used to implement the ranking algorithms. 
Although distinctions among various methods primarily arise from the underlying ranking algorithms, further differences emerge based on the characteristics of the prompts employed to implement these algorithms. 
These characteristics are not associated with the actual ranking algorithm: e.g., they may be wordings related to a role-playing strategy, or tone words. In this paper, we analysed these prompts and mapped prompt wordings into a set of components to form prompt templates that allowed us to better understand the content of these prompts. We then performed a systematic analysis of prompt variations across different types of zero-shot LLM rankers.
%Our results showed that ranking effectiveness largely varied across different instantiations of the prompt components. We found that optimal prompt wording varied across ranking methods, datasets and LLM backbones used, suggesting automatic prompt optimisation tailored to specific ranking methods and datasets, rather than manual prompt engineering, may be more appropriate to optimise ranking performance. 
Our analysis revealed that ranking effectiveness varies considerably across different implementations of prompt components. Optimal prompt wording showed variability depending on the ranking method, dataset, and LLM backbone employed, suggesting that automatic prompt optimization, tailored to specific ranking methods and datasets, may be more effective than manual prompt engineering for optimizing ranking performance. 
We make code, runs and analysis available at \url{https://github.com/ielab/zeroshot-rankers-prompt-variations}.

\clearpage
\bibliographystyle{splncs04}
\bibliography{ecir2025-prompt-variations-camera-ready}

\end{document}